\let\latexaddtocontents\addtocontents
\documentclass[a4paper,twocolumn,10pt,accepted=2024-03-21]{quantumarticle}
\let\addtocontents\latexaddtocontents\pdfoutput=1
\usepackage[utf8]{inputenc}
\usepackage[english]{babel}
\usepackage[T1]{fontenc}
\usepackage{amsmath}
\usepackage{physics}
\usepackage{hyperref}
\usepackage{url}
\usepackage{amsthm}
\usepackage{mathtools}
\usepackage{tikz}
\def\checkmark{\tikz\fill[scale=0.4](0,.35) -- (.25,0) -- (1,.7) -- (.25,.15) -- cycle;} 
\usepackage{xcolor}

\usepackage{newfloat}
\DeclareFloatingEnvironment[name={Protocol}]{enumcnt}
\graphicspath{{images/}}
\usepackage{fancyref}
\usepackage[labelfont={bf}]{caption}
\usepackage{lipsum}
\usepackage{amssymb}
\usepackage{fancyhdr,graphicx}
\usepackage[ruled,vlined]{algorithm2e}
\usepackage{multirow}
\usepackage{mathtools}
\usepackage{comment}
\usepackage{enumitem}
\usepackage{framed}
\usepackage[normalem]{ulem}
\usepackage{csquotes}

\DeclarePairedDelimiter\ceil{\lceil}{\rceil}

\theoremstyle{definition}
\newtheorem{mydef}{Definition}

\usepackage[
    backend=bibtex,
    style=phys,
  ]{biblatex}
\newcommand{\bnen}{\begin{equation}}
\newcommand{\eden}{\end{equation}}
\newcommand{\bean}{\begin{eqnarray}}
\newcommand{\eean}{\end{eqnarray}}
\newcommand{\bnsn}{\begin{subequations}}
\newcommand{\edsn}{\end{subequations}}

\newcommand{\bea}{\begin{eqnarray*}}
\newcommand{\eea}{\end{eqnarray*}}
\newcommand{\bne}{\begin{equation*}}
\newcommand{\ede}{\end{equation*}}

\usepackage{verbdef}
\verbdef{\verbperth}{ibm_perth}
\verbdef{\verbjakarta}{ibmq_jakarta}
\verbdef{\verbnairobi}{ibm_nairobi}
\verbdef{\verblagos}{ibm_lagos}
\verbdef{\verblima}{ibmq_lima}
\verbdef{\verbmanila}{ibmq_manila}
\verbdef{\verbbelem}{ibmq_belem}
\verbdef{\verbquito}{ibmq_quito}

\begin{document}

\title{Resource analysis for quantum-aided Byzantine agreement with the four-qubit singlet state}

\author{Zolt\'an Guba}
\affiliation{Department of Theoretical Physics, Institute of Physics, Budapest University of Technology and Economics, Műegyetem rkp. 3., H-1111 Budapest, Hungary}
\author{Istv\'an Finta}
\affiliation{Nokia Bell Labs}
\affiliation{\'Obuda University}
\author{\'Akos Budai}
\affiliation{Department of Theoretical Physics, Institute of Physics, Budapest University of Technology and Economics, Műegyetem rkp. 3., H-1111 Budapest, Hungary}
\affiliation{Wigner Research Centre for Physics, H-1525 Budapest, P.O.Box 49, Hungary}
\affiliation{Nokia Bell Labs}
\author{L\'or\'ant Farkas}
\affiliation{Nokia Bell Labs}
\author{Zolt\'an Zimbor\'as}
\affiliation{Wigner Research Centre for Physics, H-1525 Budapest, P.O.Box 49, Hungary}
\affiliation{E\"otv\"os University, Budapest, Hungary}
\author{Andr\'as P\'alyi}
\affiliation{Department of Theoretical Physics, Institute of Physics, Budapest University of Technology and Economics, Műegyetem rkp. 3., H-1111 Budapest, Hungary}
\affiliation{MTA-BME Quantum Dynamics and Correlations Research Group, Műegyetem rkp. 3., H-1111 Budapest, Hungary}

\maketitle

\begin{abstract}
In distributed computing, a Byzantine fault is a condition
where a component behaves inconsistently, showing different symptoms to
different components of the system.
Consensus among the correct components can be reached by appropriately crafted communication protocols even in the presence of byzantine faults. 
Quantum-aided protocols built upon distributed entangled quantum states are worth considering, as they are more resilient than traditional ones. 
Based on earlier ideas, here we establish a parameter-dependent
family of quantum-aided weak broadcast protocols.
We compute upper bounds on the failure probability of the protocol, and define and illustrate a procedure that minimizes the quantum resource requirements. 
Following earlier work demonstrating the suitability of noisy intermediate scale quantum (NISQ) devices for the study of quantum networks, we experimentally create our resource quantum state on publicly available quantum computers. 
Our work highlights important engineering aspects of the future deployment of quantum communication protocols with multi-qubit entangled states.
\end{abstract}

\tableofcontents

\section{Introduction}

In distributed computing, a Byzantine fault is a condition
where a component behaves inconsistently, showing different symptoms to
different components of the system.
To reach consensus among the correctly functioning components, 
communication protocols robust to such faults must be used.
 
An early example of such protocols was developed by Pease et al. \cite{<pease>}, where consensus is provided
if $t<n/3$, where $n$ is the number of components in the distributed system, and $t$ is the maximum number of components exhibiting Byzantine fault. 
For example, if the maximum number of faulty components is 1, then at least 4 components are needed to reach consensus. We will refer to this property of the protocol as \emph{resilience}. In fact, an important impossibility theorem (see, e.g., Sec. 5.4 of \cite{fitzi_phd}), in the framework of a standard classical model, incorporating pairwise authenticated classical communication between the components (called $\mathcal{M}_\text{aut}$ in Ref.~\cite{fitzi_phd}), restricts the existence of broadcast protocols to $t < n/3$.

In addition to resilience, communication protocols also possess other essential properties known as \textit{bit complexity} and \textit{round complexity}.  Bit complexity pertains to the number of communication cycles involved in the protocol, while round complexity refers to the length of messages transmitted during these cycles. Both of these properties play a critical role in the applicability of a communication protocol. For example, the protocol introduced in \cite{<pease>} terminates in $t+1$ rounds but requires the components to send exponentially long messages during the communication cycles. The bit complexity of early protocols \cite{<pease>,<lamport>} has since
been improved \cite{<garaymoses>,<pbft>,<raft>,<paxos>}.

Quantum-aided `weak broadcast' protocols developed in the past two decades, 
built upon distributed entangled quantum states \cite{fitzi_proceedings}
such as the four-qubit singlet state
\cite{cabello,cabstateprep}, are worth considering, as they surpass the above-mentioned classical impossibility theorem and hence offer a higher resilience:
$t$ faulty components are tolerated as long as $t < n/2$.
For example, to be resilient against $t=1$ faulty component, having $n=3$ components
is sufficient.

In this work, we build on those earlier ideas, and 
introduce a parameter-dependent family of quantum-aided weak broadcast protocols that relies on an entangled four-qubit singlet state \cite{cabello}. 
In fact, we formalize the idea described in Sec.~III. of Ref.~\cite{cabello}, which yields this protocol family. 
Then, our main contributions are as follows.

(1) \emph{Proof.} We prove the weak broadcast functionality of the protocol in a certain range of protocol parameters. The proof relies on considering the failure probability $p_f(m)$ as the function of the number $m$ of resource states, and upper bounding this failure probability exponentially, that is, proving $p_f \leq e^{-bm}$ with a certain $b>0$ as $m \to \infty$.

(2) \emph{Tight upper bound for failure probability.} We compute a tight upper bound to the failure probability of the protocol, as function of the number of quantum states used to broadcast a single classical bit.

(3) \emph{Resource optimisation.} We use the tight upper bound to perform a resource optimisation, i.e., compute the answer to these questions: (i) For a given allowed failure probability, how to set the protocol parameters to minimise the number of resource states? (ii) What is the number of required resources states?

(4) \emph{Preparation of the four-qubit singlet state.} Following earlier work demonstrating the suitability of noisy intermediate-scale quantum (NISQ) devices 
for the study of quantum networks \cite{<pathumsoot>}, we experimentally create our resource quantum 
state on the public quantum-computer prototypes of IBM and IonQ, and characterise its state fidelity.

In a spirit similar to a recent study  \cite{Taherkhani_2017}
of an alternative quantum-aided protocol \cite{<benor>}, 
our analysis illustrates multiple engineering aspects of future deployment of such protocols in practice.

The rest of the paper is structured as follows. 
In Sec.~\ref{sec:weakbroadcastprotocol},
we define a two-parameter weak broadcast protocol
based on the four-qubit singlet state. 
In Sec.~\ref{sec:securityproof}, we prove the functionality of the protocol.
In Sec.~\ref{sec:analysis}, we outline how to compute the tight upper bound for the failure probability, and discuss the resource optimisation protocol. 
In Sec.~\ref{sec:experiment}, 
we construct two four-qubit quantum circuits that can be used to generate the
singlet state, and implement those circuits on IBM's quantum computer
devices. 
We discuss our results and follow-up ideas, and draw conclusions, in 
Sec.~\ref{sec:discussion}.

\section{A two-parameter Weak Broadcast protocol}
\label{sec:weakbroadcastprotocol}

Multi-party communication protocols exist for various functionalities. 
Two examples are broadcast and weak broadcast \cite{fitzi_phd}, which are defined and compared in App.~\ref{app:broadcast}.
In short, the broadcast functionality is achieved if data is distributed consistently among correctly functioning components, whereas weak broadcast allows for an `abort' option at the receivers if the sender is an adversary. 
An important building block for more complex protocols is a weak broadcast protocol that is resilient up to one faulty component in the presence of 3 components. Following Ref.~\cite{fitzi_phd}, we will denote such protocols as WBC(3,1). A key result of Ref.~\cite{fitzi_phd} is that WBC(3,1) can be used as a
primitive, to efficiently build a broadcast protocol with $n\geq 3$ components,
resilient against any $t<n/2$ faulty components. 
This motivates us to define and study a protocol achieving WBC(3,1).

In a distributed system with 3 components, the WBC(3,1) protocol has the following goal: the sender ($S$) wants to distribute a single bit $x_S \in \{0,1\}$ of classical information to the two receivers $R_0$ and $R_1$, in such a way that there is a consensus about the bit value even if one of the three components is an active adversary.
More formally, in a WBC(3,1) protocol, the output of the receivers can take a value $\perp$ (often called `abort') besides 0 and 1, and the protocol should fulfill the conditions of `Validity' and `Consistency'. 
Validity means that if the Sender $S$ is correct (i.e., exactly follows the prescribed protocol), then the output bit values of all correct components are equal to $x_S$. 
Consistency means that if any correct component has an output $y \in \{0,1\}$, then all other correct components should have an output from $\{y,\perp\}$.
(See a more detailed description in App.~\ref{app:broadcast}.)

\begin{figure}
    \centering
    \includegraphics[scale = 0.4]{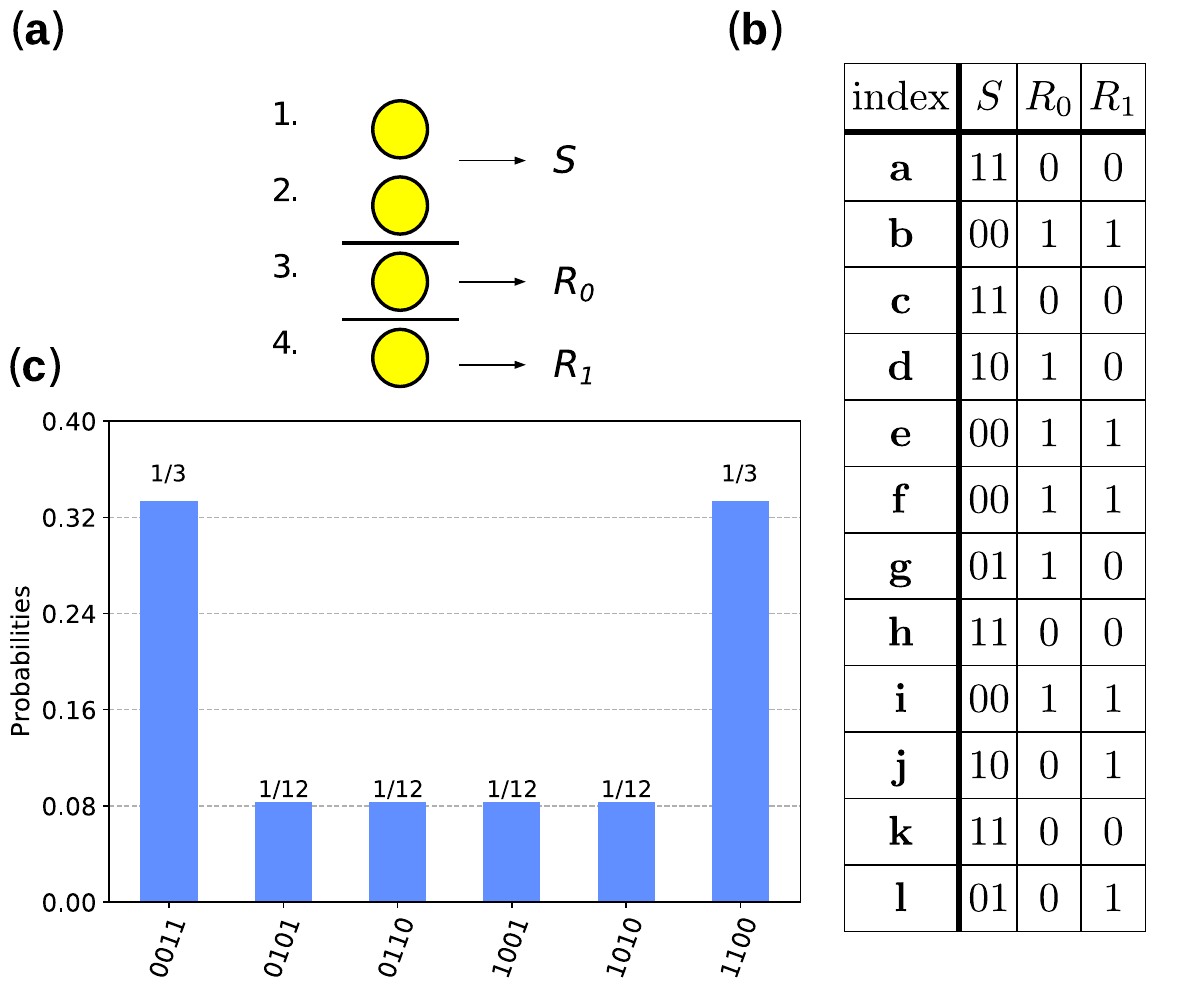}
    \caption{Using a four-qubit singlet state for
    weak broadcast. 
    (a) Yellow circles indicate the four qubits of the singlet state $\ket{\psi}$ in 
    Eq.~\eqref{eq:cabello}. 
    In the protocol, the qubits are distributed:
    qubits 1 and 2 are with the Sender ($S$), 
    qubit 3 is with $R_0$, and qubit 4 is with $R_1$.
    (b) Example of measurement outcomes from
    $m=12$ four-qubit singlet states.
    Such a table of data is called an Event.
    (c) Probability density function of the
    six possible outcomes when the four-qubit 
    singlet state $\ket{\psi}$ is measured.}
    \label{fig:Cabstate_distribution}
\end{figure}

The two-parameter WBC(3,1) protocol we define here is derived from similar protocols
in earlier works \cite{fitzi_proceedings,cabello,cabstateprep}.
It prescribes communication between three components ($S$, $R_0$, $R_1$) to achieve consensus in the sense of WBC(3,1) described above.
The quantum resource of this protocol is a four-qubit singlet state \cite{cabello}:
\begin{align}
\label{eq:cabello}
\begin{split}
    \ket{\psi}   =\frac{1}{2 \sqrt{3}} (2\ket{0011}-\ket{0101}- \ket{0110}-
    \\ 
    \ket{1010}-\ket{1001}+2\ket{1100}).
\end{split}
\end{align}
In fact, we assume that to achieve WBC(3,1) with a single data bit, 
the parties have distributed $m \in \mathbb{Z}^+$ four-qubit singlet states
through public quantum channels such that the first two qubits are distributed to $S$, the third qubit is distributed to $R_0$,
and the last qubit is distributed to $R_1$, as shown in Fig.~\ref{fig:Cabstate_distribution}a. (See Sec.~\ref{sec:discussion} for a brief discussion on the distribution.)
In rough terms, the more four-qubit singlet states are used to broadcast one bit, the higher the probability of success; hence the integer $m$ is a key quantity in assessing the performance of the protocol.

Having $m$ singlet states distributed, the components perform measurements in the 
computational basis, each obtaining its own random list of bit values:
$S$ obtains $m$ bit-pairs, while $R_0$ and $R_1$ obtain $m$ bits each. 
An example for such a collection of random correlated bitstrings
for $m=12$ is shown in Fig.$~\ref{fig:Cabstate_distribution}$b,
and the probability density function 
of the six possible outcomes of four-bit strings, derived straightforwardly from Eq.~\eqref{eq:cabello},
is shown in Fig.$~\ref{fig:Cabstate_distribution}$c.
We will refer to the six possible four-long bitstring outcomes
as \emph{Elementary Events}, and we will refer to 
a particular instance of this collection of random data, as shown in Fig.~\ref{fig:Cabstate_distribution}b, as an \emph{Event}.
In what follows, we will use $Q_j[\alpha]$ to denote the
measurement result of component $j$ in its row $\alpha$,
where $j \in (S,R_0,R_1) \equiv (S,0,1)$
and
$\alpha \in (1,2,\dots, m) \equiv (\text{a},\text{b} \dots)$.
For example, in Fig.~\ref{fig:Cabstate_distribution}b, we see $Q_S[\text{d}] = 10$, $Q_0[\text{e}] = 1$, etc.

The Weak Broadcast protocol presented below
has two real-valued parameters:
$0 < \mu < 1/3$ and $1/2 < \lambda < 1$.
We anticipate that the protocol 
works well if $\mu$ and $\lambda$ are
close to their upper bound. 

The Weak Broadcast protocol we propose is as follows.
In all steps, the classical channels are assumed to be
authenticated, and all the channels, 
the source of quantum states, and the measurements, 
are assumed to be perfect. 

\begin{enumerate}
    \item 
    \emph{Invocation Phase.}
    $S$ sends the data bit ($x_S$) to be broadcasted to the other components.
    We denote the bits of the components $R_0$ and $R_1$ received 
    as $x_0 = x_S$ and $x_1 = x_S$. 
    Now the Sender measures all its qubits ($2$ qubits in each distributed state) in the computational basis.
    For each qubit pair of $S$, if both measurements corresponding to
    state index $\alpha$ 
    (see Fig.~\ref{fig:Cabstate_distribution}b)
    yielded $x_S$, then 
    $S$ adds index $\alpha$ to its \emph{check set} $\sigma_S$.
    After assembling $\sigma_S$, 
    this check set is sent to both receivers.
    At this point, both receivers hold a bit value $x_j$ and a set $\sigma_j$ from the Sender. 
    The Sender also sets its output to $y_S = x_S$.
    In the example of Fig.~\ref{fig:Cabstate_distribution}b, 
    and assuming $x_S = 0$, the check set is $\sigma_S = \{\mathrm{b},\mathrm{e},\mathrm{f},\mathrm{i}\}$.

    \item 
    \emph{Check Phase.}
    Now both $R_0$ and $R_1$ check the consistency of their data received from $S$. For this, $R_j$ measures all of its qubits in the check set $\sigma_j$, and if all the results differ from $x_j$ 
    (\emph{Consistency Condition}), 
    and also the check set is large enough
    \emph{(Length Condition)},
    then it accepts the message $x_j$. 
    For $R_0$, this implies that it chooses its
    output to be $y_0 = x_0$.
    For $R_1$, it chooses the value of an intermediate 
    variable $\tilde{y}_1 = x_1$.
    The check set is large enough if the number
    of its elements is at least
    $T \equiv \lceil \mu \cdot m \rceil$, where $0<\mu<1/3$.
    If $R_0$ finds that any of the two conditions is violated, then it sets its output to `abort', $y_0 = \perp$.
    If $R_1$ finds that any of the two conditions is violated, then it sets its intermediate value to `abort', $\tilde{y}_1 = \perp$.

    \item 
    \emph{Cross-calling Phase.}
    $R_0$ sends to $R_1$ its output value $y_0$ 
    and the check set $\sigma_0$ it received from $S$.
    $R_1$ receives these as $y_{01}$ and $\rho_{01}$,
    respectively.

    \item 
    \emph{Cross-check Phase.}
    $R_1$ evaluates the following three conditions
    and if all of them are true, then it outputs the value
    received from $R_0$, that is, $y_1 = y_{01}$; 
    otherwise it outputs its intermediate value, $y_1 = \tilde{y}_1$.
    The three conditions: 
    (i) Confusion Condition: $y_{01}$ is different from 
    $\tilde{y}_1$, and none of them is $\perp$.
    (ii) Length Condition: 
    The size of the check set $\rho_{01}$ is at least $T \equiv \lceil \mu \cdot m \rceil$, 
    similarly to the Length Condition of the Check Phase above. 
    (iii) Consistency Condition:
    For a large fraction of the indices in $\rho_{01}$, 
    $R_1$ measures the bit value opposite to $y_{01}$. 
    The tolerance parameter defining this fraction is $\lambda$, 
    assumed to fulfill $1/2<\lambda <1$;
    that is, this condition is true if the number of indices in $\rho_{01}$
    where $R_1$ measures the bit opposite to $y_{01}$ is greater or equal to 
    $\lambda T + |\rho_{01}| - T$.
    Note that this Consistency Condition is less stringent than the 
    one in the Check Phase.

\label{lst:fitzicabello_text}
\end{enumerate}

\begin{figure}[t]
    \centering
\includegraphics[scale = 0.5]{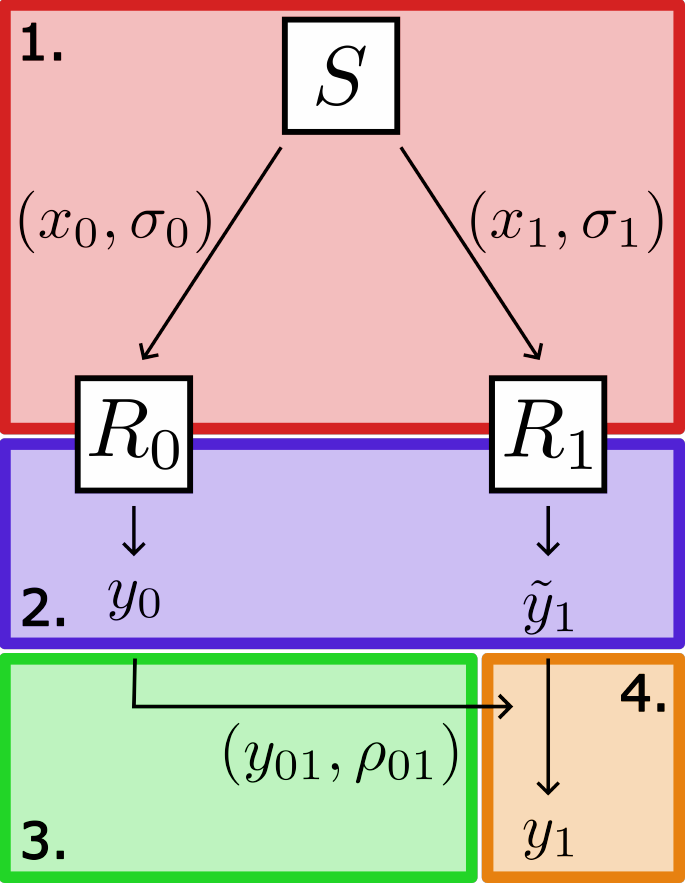}
\caption{Flowchart showing the classical communication part of the protocol. 
Arrows with symbols in brakets show the use of the classical communication channels.
Further arrows denote the decision-making processes of the components.}
\label{fig:prot_flowchart}
\end{figure}

We also provide a more compact and more formal
definition of the protocol:

\begin{enumerate}

    \item $S \rightarrow R_{0}$, $R_{1}$: $x_{S}, \sigma_{S}=\{\alpha \in\{1, \ldots, m\}: Q_{S}[\alpha]=x_{S} x_{S}\}, R_{j}$ receive $\left\{x_{j}, \sigma_{j}\right\}$
$S: y_{S}=x_{S}$
    
    \item $R_{0}:$  if $(| \sigma_{0} | \geq T)$ $\wedge$ $( \{\alpha \in \sigma_{0}: Q_{0}[\alpha]=x_{0}\}=\emptyset)$ then $y_{0}=x_{0}$ else $y_{0}=\perp$ fi  
    
    $R_{1}:$  if $(| \sigma_{1} | \geq T)$ $\wedge$ $( \{\alpha \in \sigma_{1}: Q_{1}[\alpha]=x_{1}\}=\emptyset)$ then $\tilde{y}_{1}=x_{1}$ else $\tilde{y}_{1}=\perp$ fi

    \item  $R_{0} \rightarrow R_{1}: y_{0}, \rho_{0}= \sigma_0 ; R_{1}:$ receives $ (y_{01}, \rho_{01} )$
    
    \item  $R_1$: if $ ( \perp \neq y_{01} \neq y_{1} \neq \perp ) \wedge ( |\rho_{01}  | \geq T ) \wedge$
    
    $\wedge ( |\{\alpha \in \rho_{01} : Q_{1}[\alpha]=1-y_{01} \}| \geq  \lambda T + |\rho_{01} | - T)$
then $y_{1}:=y_{01}$ else $y_1 = \tilde{y}_1$ fi

\captionof{enumcnt}{Formal description of the Weak Broadcast protocol.}
\label{lst:fitzicabello}
\end{enumerate}
The flowchart of the protocol can be seen in Fig.~\ref{fig:prot_flowchart}.

In what follows, we will use the label `Weak Broadcast protocol' or simply `Weak Broadcast'
to denote the above protocol. 
The description above shows how the correct components behave.
Our goal is to show that this 
protocol is secure for at most one faulty component in a certain range of the values of the parameters $\mu$ and $\lambda$. 
Roughly speaking (see below for the theorem), 
by security \cite{fitzi_phd} we mean that the \emph{failure probability} of the protocol, that is, the probability of not achieving weak broadcast,
converges to zero as the number $m$ of four-qubit singlet states is increased to infinity. 
Even though the Weak Broadcast protocol itself 
follows straightforwardly from earlier works
\cite{cabello,cabstateprep}, the security theorem
and its proof has not been published before, to our knowledge.

We point out an important property of the above Weak Broadcast protocol, which is shared with similar earlier protocols \cite{fitzi_phd,fitzi_proceedings,<fitzi_2001>,cabstateprep}. There is an asymmetry between the two Receivers $R_0$ and $R_1$. In contrast to $R_0$, $R_1$ never sends messages, thus it has no impact on the output of the other components. In particular, if $R_1$ is the faulty component,
then the other two components will reach consensus and hence weak broadcast is guaranteed. 
Therefore, from now on we do not consider
the case when $R_1$ is faulty; we restrict our attention
to three other adversary configurations, which we will refer to as 'no faulty', `S faulty', and `$R_0$ faulty'.

Two key advantages of this protocol, with respect to other quantum-assisted protocols, are as follows.
(1) This protocol is based on qubits, in contrast to earlier ones based on multi-level qudits, e.g., \cite{<fitzi_2001>,<benor>}.
At this early stage of quantum communication technology, most of the experimental effort is focused on qubit-based protocols, hence we consider this as a significant advantage favoring the protocol we study. 
(2) The resource state of Eq.~\eqref{eq:cabello} is readily available experimentally \cite{cabstateprep} in terms of entangled, polarisation-encoded photonic qubits, via spontaneous parametric down-conversion.

\section{Security of Weak Broadcast}
\label{sec:securityproof}

The Weak Broadcast protocol defined in Sec.~\ref{sec:weakbroadcastprotocol}
is based on Refs.~ 
\cite{fitzi_proceedings,fitzi_phd,cabello,cabstateprep}.
Reference \cite{fitzi_phd} provides a security proof of the 
weak broadcast protocol proposed therein.
That protocol uses entangled three-qutrit states.
In this work, we apply elements from there to construct
a security proof for the above-defined Weak Broadcast 
protocol, which is based on four-qubit singlet states. 
Preliminaries and the theorem are presented in this section; 
for the proof we refer to App.~\ref{app:securityproof}.

We start by specifying our framework.
We assume that there is a complete network of pairwise authenticated 
classical channels 
among the three components, $S$, $R_0$ and $R_1$. 
We also assume that this classical network is synchronous.
In particular, 
(1) before the protocol, the components have already agreed on a common point in time when the protocol is to be started,
and 
(2) all components operate according to a global clock and every message
sent during a clock cycle 
(our protocol consists of 2 clock cycles)
will arrive to the receiver by the end of the cycle.
Recall also that we consider three \emph{adversary configurations}:
`no faulty', `$S$ faulty', and `$R_0$ faulty'.

Here, \emph{adversary} or \emph{faulty component} means
an \emph{active adversary}, which might act as a properly functioning
component, but it might try to undermine the protocol by sending
confusing information to the other components.
In fact, we assume a conscious adversary
following a rational strategy.
We also assume that the adversary has only local information.
In other words, each random bit generated
from the four-qubit singlet states by a local measurement,
see Fig.~\ref{fig:Cabstate_distribution}b, is known only for 
the component that has performed the measurement.

We also assume that there is a quantum source that distributes the
four-qubit singlet states to the components as shown in 
Fig.~\ref{fig:Cabstate_distribution}.
This distribution happens before the measurements required for the
communication steps of the Weak Broadcast protocol are performed. 

The protocol we describe above is a probabilistic one, since it is based on measurements of a quantum state that is the superposition of the measurement basis states.
The number $m$ of four-qubit singlet states consumed in the protocol is a parameter of the protocol, and the probability of success and failure, defined through the conditions of Validity and Consistency above (see also App.~\ref{app:broadcast} for more details) depends on $m$.
We denote the \emph{failure probability} with $p_f(m)$.
Furthermore, we call the protocol \emph{secure}, if the failure probability converges to zero ($p_f(m) \to 0$) as the number of consumed states tends to infinity ($m\to \infty)$.

After these preliminaries, we state the theorem 
claiming the security of Weak Broadcast:

{\bf Theorem:} For any parameter pairs $(\mu,\lambda)$ 
fulfilling 
\begin{subequations}
\label{eq:conditionsoftheorem}
\begin{eqnarray}
\label{eq:conditionmu}
2/9 &<& \mu < 1/3, \mbox{ and} \\
\frac{2+9\mu}{18\mu} &<& \lambda < 1,
\label{eq:conditionlambda}
\end{eqnarray}
\end{subequations}
the failure probability $p_f(m)$ of the Weak Broadcast protocol, defined as Protocol \ref{lst:fitzicabello} above, converges to zero as $m \to \infty$. 
In particular, in all three adversary configurations, the failure probability 
is upper-bounded as
\begin{equation}
    p_f(m) \leq a e^{-b m},
\end{equation}
with $a,b>0$.

\begin{figure}
    \centering
    \includegraphics[scale = 0.6]{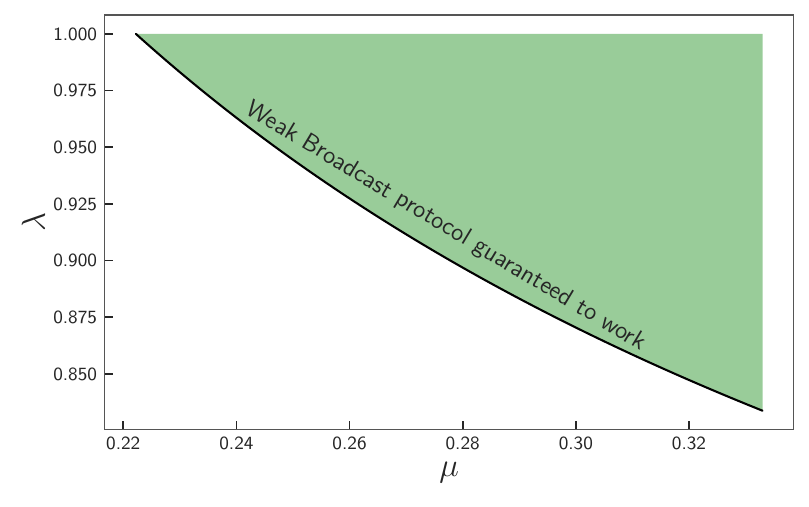}
    \caption{
    Parameter region where the Weak Broadcast 
    protocol (Protocol \ref{lst:fitzicabello}) is guaranteed to work. 
    I.e., for parameter pairs $\mu,\lambda$
    in the green region (see Eq.~\eqref{eq:conditionlambda}) 
    the failure probability of the 
    Weak Broadcast protocol 
    can be upper-bounded
    as $p_f(m) \leq a e^{-b m}$ with $a,b>0$.
    }
    \label{fig:R0FaultyCompleteCondition}
\end{figure}

The proof is found in App.~\ref{app:securityproof}.
The parameter range where our proof guarantees security is shown in Fig.~\ref{fig:R0FaultyCompleteCondition}.
To enable the proof, we introduce the optimal adversary strategies in App.~\ref{app:adversarystrategies} and prove their optimality in App.~\ref{app:proofsofoptimality}.
Then, the proof in App.~\ref{app:securityproof} is based on the knowledge of the optimal adversary strategies.

In the spirit of Refs.~\cite{fitzi_phd,fitzi_proceedings}, we make a remark on the complexity of the protocol: 
Let $\mathcal{R}_\mathrm{q}$ and $\mathcal{B}_\mathrm{q}$ be the round and bit complexities of distributing and measuring a four-qubit singlet state. Then, the Weak Broadcast protocol with a failure probability of at most $0 < \epsilon \ll 1$ requires a round complexity of $\mathcal{R} = \mathcal{R}_q+2$, and a bit complexitiy of $\mathcal{B} = \mathcal{O}(\log(1/\epsilon)) \mathcal{B}_q$.

A natural question is if using secure classical channels
instead of authenticated ones brings any advantage.
Here we reason that the answer is no. 
If the Sender is the faulty component, 
then even if it can eavesdrop on the communication between
$R_0$ and $R_1$, that has no effect on the outcome of the 
protocol. 
Furthermore, if $R_0$ is the faulty component, then 
$S$ is correct, hence $S$ sends the very same information to
both receivers.
Therefore, $R_0$ eavesdropping on the communication between
$S$ and $R_1$ does not gain extra information for $R_0$. 
These imply that authenticated classical 
channels where eavesdropping is possible
are just as useful in this case as secure channels.

\section{Resource optimization of the Weak Broadcast protocol}
\label{sec:analysis}

In this section, our first goal is to calculate a tight upper bound for the failure probability $p_f(m)$ of the Weak Broadcast protocol.
This tight upper bound will be used to estimate the resource requirements of the protocol. 
In particular, we will estimate the number $m$ 
of singlet states that are required to achieve 
weak broadcast with a small failure probability, say, 5\%. 
This analysis is similar in spirit to Ref.~ \cite{Taherkhani_2017}, which was carried out
for a different quantum-aided multi-party communication
protocol \cite{<benor>}.

First, we derive the tight upper bound of the failure probability of the Weak Broadcast protocol as function of $\mu$ and $\lambda$.
We obtain analytical results, and cross-check those against Monte-Carlo simulations. 
Then, we optimize the protocol in the $\mu$-$\lambda$ parameter plane. 
That is, we identify
the parameter values that require the least resources, i.e., the smallest number of four-qubit singlet states such that the failure probability is guaranteed to be below the desired failure probability threshold.

\subsection{Tight upper bound for the failure probability -- analytical results}
\label{anexpressions}

We analyse the resource requirement of the Weak Broadcast protocol is as follows. 
We can fix the parameter values $\mu$ and $\lambda$, and the \emph{target failure threshold} $p_{f,t}$;
e.g., $\mu =0.272$, $\lambda = 0.94$, and $p_{f,t} = 0.05$.
The target failure threshold $p_{t,f} = 0.05$ means that the user can tolerate at most 5\% failure probability. 
This is an arbitrarily chosen small number here; we envision that in future applications of such probabilistic consensus protocols, the user must specify such a desired level of tolerance against statistical errors. The method we describe and exemplify here is applicable irrespective of the specific value of $p_{f,t}$.
In this setting, the resource requirement of the protocol is characterized by the minimal number $m_\text{min}(\mu,\lambda,p_{f,t})  \in \mathbb{Z}^+$ of four-qubit singlet states required to suppress the failure probability $p_f$ below the threshold $p_{f,t}$.

A natural way to compute $m_\text{min}$ would be to identify the optimal adversary strategies for the $S$ faulty and $R_0$ faulty configurations, i.e. those strategies that maximize the failure probability.
We do not complete this task here. 
Instead, we identify optimal \emph{incomplete} strategies, which are incomplete in the sense that they provide a description of the adversary's behavior for certain Events but not for others. 
Using these optimal incomplete strategies, we can compute tight upper bounds on the failure probabilities $p_f(m)$, and can use those upper bounds to compute an upper bound $m_{\text{min},\uparrow}$ of the minimal resource requirement. 

For the no faulty configuration, we derive the exact result $p_f^\text{(nf)}$ for the failure probability shown as Eq.~\eqref{eq:pfnofaulty}.
The optimal incomplete adversary strategies for the $S$ faulty and $R_0$ faulty configurations are described in App.~\ref{app:adversarystrategies} and App.~\ref{app:proofsofoptimality}. 
The corresponding failure-probability upper bounds $p_{f,\uparrow}^{(S)}$ and
$p_{f,\uparrow}^{(R)}$ are derived in App.~\ref{app:failureprobabilities}, and are  shown in Eq.~\eqref{eq:ps_upper} and  Eq.~\eqref{equation:pr0_upper}, respectively.

\begin{figure}[t]
    \centering
    \includegraphics[scale = 0.9]{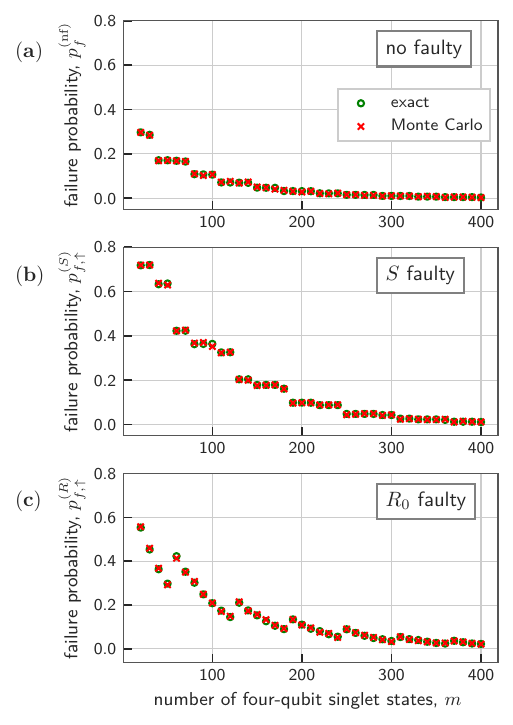}
    \caption{Failure probability of the Weak Broadcast protocol for the three adversary configurations.
    (a) `No faulty' configuration. Exact results (green circles) and Monte-Carlo results (red crosses) for the failure probability.
    (b)`$S$ faulty configuration.
    (c) `$R_0$ faulty' configuration.
    In (b) and (c), the failure-probability upper bound is shown.
    Parameters (all panels): $\mu = 0.272$ and $\lambda = 0.94$.
    Simulation results are obtained from $N=10,000$ random Events for each $m$.
    (Fig.~\ref{fig:Cabstate_distribution}b shows a single random Event for $m=12$.)
    \label{fig:theoryandsimulation}}
\end{figure}

To evaluate these analytical formulas, we carry out numerical summation. 
The results, for a fixed value of protocol parameters, are shown in the three panels of Fig.~\ref{fig:theoryandsimulation}, corresponding to the
three adversary configurations. See caption for parameter values.
All three panels show that the failure probability $p_f$ exhibits a 
decreasing trend as we increase the number of 
four-qubit singlet states, as expected.
From the data, we can read off the minimum number of states
required to suppress the upper bound of the failure probability below $p_{f,t} = 0.05$:
these are 143, 246 and 280, respectively, 
for the three adversary configurations. 
Our upper bound $m_{\text{min},\uparrow}$ of the minimal resource requirement of this protocol for $p_{f,t} = 0.05$ is the maximum of those
three integers, that is, $m_{\text{min},\uparrow} = 280$.

\subsection{Tight upper bound for the failure probability - Monte-Carlo simulations}
\label{sec:montecarlo}

To test our analytical calculations described in the previous subsection, here we use Monte-Carlo simulations to compute the failure probability $p^{\text{(nf)}}_f$ and the failure-probability upper bounds $p^{(S)}_{f,\uparrow}$ and $p^{(R)}_{f,\uparrow}$.

In these simulations, we randomly generate, on a classical computer, the lists of correlated 4-bit strings measured by the components (i.e., the Events, see Fig.~\ref{fig:Cabstate_distribution}b).
For the no faulty configuration, we evaluate for each random Event if the protocol leads to failure or weak broadcast.
Then, we express the failure probability as the ratio $N_f/N$ of the number $N_f$ of failures and the number $N$ of random Events. 
The resulting failure probability as function of $m$ is shown as the red crosses in Fig.~\ref{fig:theoryandsimulation}a.
Data corresponds to $N=10,000$ random Events. 

For the $S$ faulty and $R_0$ faulty adversary configurations, our simulation corresponds to the upper-bound calculation described in the previous subsection. 
For example, for the $S$ faulty configuration, we use the optimal incomplete strategy described in App.~\ref{app:adversarystrategies}, and do the following steps. 
(1) We generate a random Event. 
(2) We check if the random Event is in the domain of the optimal incomplete strategy. 
(3a) If the Event is not in the domain, then we assume failure. 
(3b) If the Event is in the domain, then we evaluate the protocol such that $S$ is acting according to its adversary strategy; this leads to either failure or weak broadcast.
(4) After repeating these steps for $N$ different random Events, we express the failure-probability upper bound as the ratio $N_f/N$ of the number $N_f$ of failures and the number $N$ of random Events. 
These results are shown as the red crosses in Fig.~\ref{fig:theoryandsimulation}b and c, for the $S$ faulty and $R_0$ faulty configurations, respectively.
Data corresponds to $N=10,000$ random Events.

In Fig.~\ref{fig:theoryandsimulation}a, b, and c, the results of the Monte-Carlo simulations (red crosses) are in excellent agreement with the exact results (green circles).
We interpret the small differences between the Monte-Carlo result and the exact result as statistical fluctuations of the Monte-Carlo result. 
For example, estimating the failure probability $p=0.2$ from $N=10,000$ samples has a statistical error of
$\sqrt{p(1-p)/N} \approx 0.4$ \%.

\subsection{Optimization in the parameter space}
\label{sec:optimization}

In the previous subsections, we computed the resource requirements
of the Weak Broadcast protocol for fixed values of the parameters $\mu$ and $\lambda$,
and for a fixed value of the target failure threshold $p_{f,t}$.
A natural next step is the optimisation of the protocol: finding the optimal parameter values $\lambda$ and $\mu$, which minimize the resource requirements, such that it is guaranteed that the failure probability is kept below the target failure threshold $p_{f,t}$.
Even though we do not know the failure probabilities of the best adversary strategies, we do know tight upper bounds for the failure probability, using which we can carry out the optimization.
In particular, we will minimize the function $m_{\text{min},\uparrow}(\mu,\lambda,p_{f,t})$
in the $(\mu,\lambda)$ parameter plane, with the specific value of the target failure threshold $p_{f,t} = 0.05$. 
We expect that future deployment of such communication protocols will require similar optimization procedures.

\begin{figure}[t]
    \centering
    \includegraphics[scale = 0.4]{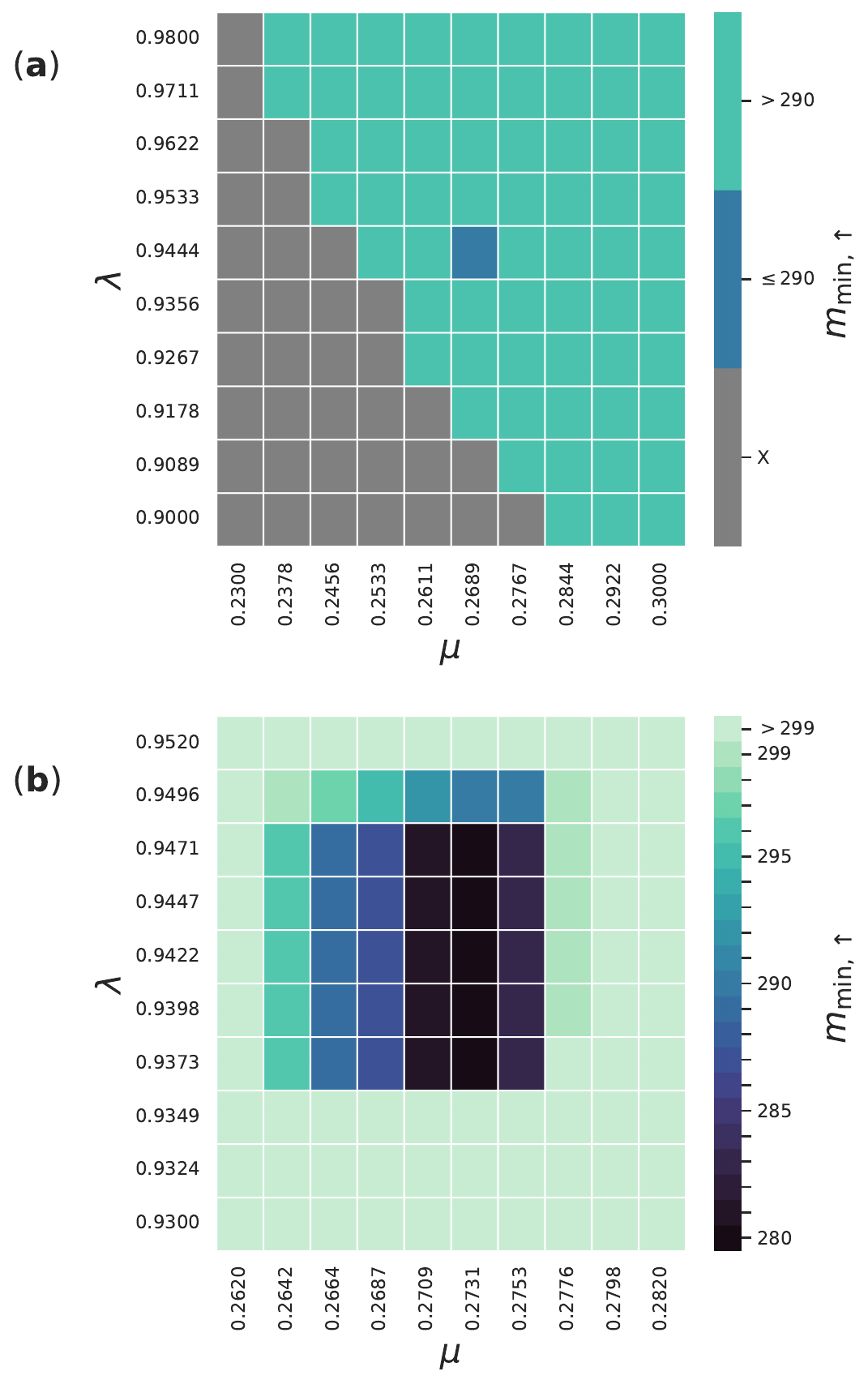}
    \caption{Optimization of the Weak Broadcast 
    protocol in the $(\mu,\lambda)$ parameter space,
    for a fixed target failure threshold.
    Both panels indicate the resource requirement, i.e., 
    number of four-qubit singlet states, 
    to achieve failure probability below the
    target failure threshold $p_\text{th} = 0.05$. 
    Grey boxes indicate parameter points that are outside of the 
    guaranteed range of the protocol (i.e., are outside of the green range
    in Fig.~\ref{fig:R0FaultyCompleteCondition}).
    (a) Rough optimization.
    Dark pixels indicate the more economical 
    region in the parameter space.
    (b) Fine optimization: zoom-in of (a).
    Dark pixels (`sweet rectangle') indicate
    the most economical region 
    in the parameter space.}
    \label{fig:parspace}
\end{figure}

We perform this optimization in two steps, as follows.
We use the analytical formulas discussed in section \ref{anexpressions}.
For a rough optimization in the $(\mu,\lambda)$ parameter space, 
we consider the rectangle in Fig.~\ref{fig:parspace}a, and the grid of
parameter pairs shown there.
For each point (box) of the grid, we evaluate the failure-probability upper bound $p_{\uparrow}(m)$ for 
$m=290,300$, and color the box as blue or green, respectively, according to the minimal $m$ within this set where $p_{f,\uparrow}<p_{f,t} = 0.05$.
This procedure results in the pattern in Fig.~\ref{fig:parspace}a, 
which suggest a finer optimization focusing on the area
of the dark blue rectangle of Fig.~\ref{fig:parspace}a. 
Note that grey boxes correspond to parameter values $\mu$, $\lambda$ lying outside the green zone of Fig.~\ref{fig:R0FaultyCompleteCondition}.

The results of this finer optimization are shown in 
Fig.~\ref{fig:parspace}b. 
The procedure is analogous to the rough optimization, 
with the difference that here the set of $m$ values is
$m = 270, 271, \dots, 300$. 
Figure ~\ref{fig:parspace}b reveals a `sweet rectangle', a rectangle
of points where using 280 four-qubit singlet states is sufficient
to achieve weak broadcast with at most 5 \% error. 
We emphasize that for an error threshold different from 5 \%, this optimization
must be repeated, and might yield different optimal values for the parameters
$\mu$ and $\lambda$.

\subsection{Robustness of the optimized protocol against physical errors}
\label{subsec:robustness}

In the previous subsections, we assumed that the Weak Broadcast protocol functions perfectly, i.e., we neglected physical errors.  
Such errors are expected to increase the failure probability of the protocol; in this subsection, we estimate this increase.

To simplify the error analysis, we use a phenomenological error model, based on three key assumptions: 
(1) We assume that the process of generating and measuring the distributed four-qubit singlet state is affected by physical errors, and the effect of those is described by a modified measurement statistics over the 16 computational basis states, i.e., a noisy measurement statistics that is slightly different from the ideal one defined by Eq.~\eqref{eq:cabello}.
(2) We assume that a slight rearrangement of the measurement statistics among the six `useful' computational basis states (those involved in Eq.~\eqref{eq:cabello}) is harmless, but `leakage' of the noisy statistics to the 10 remaining `useless' computational basis states is harmful.
In particular, we call the probability weight of the noisy statistics on the useless states the \emph{noise strength} $q$.
(3) We assume a pessimistic scenario, where even a single leakage event implies the failure of the protocol.

According to these assumptions, the failure probability of the protocol in the noisy case reads
\begin{equation}
    \label{eq:noisy}
    p_f^\mathrm{n} = \left( 1 - P \right) p_f + P,
\end{equation}
where $P$ is the probability that there is at least one leakage event, which is expressed as: 
\begin{equation}
    \label{eq:leakage}
    P = \sum_{i=1}^m \binom{m}{i} q^i (1-q)^{m-i} = 1- (1-q)^m.
\end{equation}

In Sec.~\ref{sec:optimization} we concluded that in order to reach a target failure probability $p_{f,t} = 5$\%, consuming $m \approx 300$ four-qubit singlet states is sufficient. 
Consider now the situation that the user of the protocol can accommodate an extra 1\% failure probability due to physical errors not taken into account so far.
Based on Eq.~\eqref{eq:leakage}, we find that the extra failure probability is kept below 1\% as long as the noise strength fulfills $q < 3.3 \times 10^{-5}$.

\subsection{Achievable failure probability in the presence of physical errors}

In the ideal, noiseless case, for a fixed parameter point $(\mu,\lambda)$ in the green region of Fig.~\ref{fig:R0FaultyCompleteCondition}, the failure probability $p_f$ of the Weak Broadcast protocol tends to zero as $m$ tends to infinity.
This is no longer the case in the noisy case described by Eqs.~\eqref{eq:noisy} and \eqref{eq:leakage}, since $P(q,m)$ expressed in Eq.~\eqref{eq:leakage} is a monotonically increasing function of $m$ for $0<q<1$.

However, since the estimate $p_f^\textrm{n}$ for the noisy failure probability, see Eq.~\eqref{eq:noisy}, is a sum of a decreasing and an increasing function of $m$ (for a fixed $q$), it is reasonable to expect the existence of an optimal $m$ where $p_f^\textrm{n}$ is minimised.
For example, $\mu = 0.272$, $\lambda = 0.94$, and $q = 10^{-4}$, we have numerically found a minimal failure probability of $p_{f}^\textrm{n} = 5.5$\%, corresponding to an optimal number of four-qubit singlet states of $m=423$.
It is straightforward to incorporate this simple noise description into the optimisation process introduced in Sec.~\ref{sec:optimization}.

To summarize, in this section, we have outlined and implemented
resource optimization procedures for the Weak Broadcast protocol. 
This illustrates the engineering aspects of the deployment
of future quantum-aided distributed systems. 
Furthermore, we provided a simple noise analysis, which  illustrated that a finite noise implies a nonzero achievable failure probability for the Weak Broadcast protocol. 
In other words: the noise strength has to be kept below an upper bound determined by the target failure probability.

\section{Four-qubit singlet state on IBM Q and IonQ devices}
\label{sec:experiment}

An important goal in the field of quantum communication is
to advance the hardware technology and thereby 
to enable the practical deployment 
of quantum-aided multi-party communication \cite{cabstateprep,Wehner,Pompili}.
With this goal in mind, we studied the preparation of the resource
states of the Weak Broadcast protocol, i.e., four-qubit singlet states, 
on NISQ hardware \cite{preskill2018quantum}, namely using superconducting qubits of IBM Q \cite{ibm} and trapped ion qubits of IonQ \cite{monroe2021ionq}.

Both superconducting and trapped ion qubits have already been shown to be promising candidates 
for implementing quantum networks \cite{<pathumsoot>}.
Quantum network functionalities using 
superconducting qubits housed in spatially
separated cryogenic systems have
been experimentally demonstrated recently \cite{Magnard,Storz}, while scalable remote entanglement between trapped ions has been achieved via photon and phonon interactions \cite{hucul2015modular,Krutyanskiy,Stephenson}, raising the hope that entangled states prepared
on-chip on quantum processors can be used
as resources in quantum networks.

In fact, one way to prepare a distributed four-qubit singlet state  is to prepare it locally, and then make use of Bell pairs pre-shared via quantum links \cite{Magnard,Storz,Krutyanskiy,Stephenson} to perform quantum teleportation to distribute the four-qubit state. For such a scheme, which, to our knowledge, has not yet been realised, an analysis of local state-preparation fidelity provides an upper bound of the fidelity of the distributed state.
This motivates us to do experiments on state-of-the-art \emph{quantum computing} hardware, composed of locally connected qubits, and test the precision of state preparation of the four-qubit singlet state.

In the rest of this section, we identify two quantum circuits that prepare the four-qubit singlet state and report the results of the state preparation experiments using those circuits.
To characterize the quality of our experimental state preparation, we use three quantities.
The first one is the \emph{classical fidelity} $F_c$, that is the fidelity between (i) the 
probability distribution $P_\text{exp}$ of bitstrings 
obtained experimentally by doing computational-basis
measurements, 
and (ii) the ideal probability distribution 
$P_\text{id}$
of measured bitstrings, following from 
Eq.~\eqref{eq:cabello} and 
shown in Fig.~\ref{fig:Cabstate_distribution}c:
\begin{equation}
    \label{eq:classicalfidelity}
    F_c =  \left( \sum_{s = (0000)}^{(1111)}
    \sqrt{P_\text{exp}(s) P_\text{id}(s)} \right) ^2.
\end{equation}

The second figure of merit for state preparation
is the \emph{quantum state fidelity} $F_q$.
We obtain this by performing quantum state
tomography at the end of the circuit, 
reconstructing the density matrix $\rho_\text{exp}$
from the experimental data, 
and evaluating the fidelity between this
state and the ideal
four-qubit singlet state 
$\rho_\text{id} = \ket{\psi}\bra{\psi}$:
\begin{equation}
    \label{eq:quantumfidelity}
    F_q = \left(  \mathrm{Tr}\sqrt{ \rho_\text{id}^{1/2} \rho_\text{exp}
    \rho_\text{id}^{1/2}} \right) ^2.
\end{equation}

The third figure of merit we evaluate is the `noise strength' $q$ defined in Sec.~\ref{subsec:robustness}.

\subsection{Loop Circuit}

First, we need to identify a quantum circuit that
prepares the four-qubit singlet state. 
We rely on 
CNOT as the two-qubit gate.
On today's noisy devices, CNOT is `expensive' in the sense
that its gate error is greater than the single-qubit 
gate errors. 
Therefore, a second target for our circuit search is to minimize the number of CNOTs.
A third aspect is related to the incomplete connectivity
of IBM Q devices: we pay special attention to devices that host a chain of four qubits with linear connectivity.

\begin{figure}[t]
    \centering
    \includegraphics[scale = 0.5]{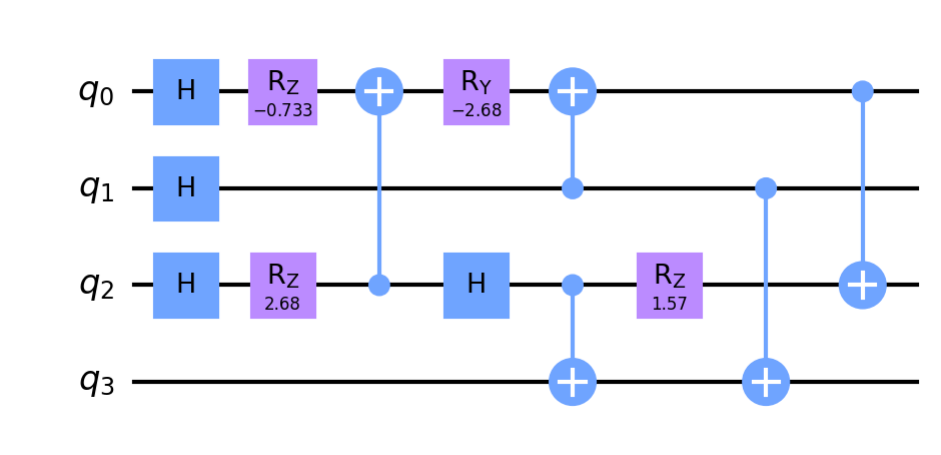}
    \caption{Loop Circuit, a 5-CNOT circuit that
    prepares the four-qubit singlet state
    of Eq.~\eqref{eq:cabello}. The decimal fractions present in the circuit map to -0.73304, 2.67908 and 1.5708.}
    \label{fig:circA}
\end{figure}

A four-qubit circuit that prepares 
the four-qubit singlet state $\ket{\psi}$, which we call the Loop Circuit, is depicted in Fig.~\ref{fig:circA}.
It includes 5 CNOT gates that define a loop topology among the four qubits. 
We found the Loop Circuit using a combination
of random circuit search and gradient descent:
We assembled random circuits using CNOTs and
single-qubit gates from the set
$\{I,X,T,S,H\}$; searching through approximately $10^7$ such
instances, we selected the one whose
output state had the maximal overlap with the 
four-qubit singlet state.
Then, for the selected instance, we replaced all the one-qubit gates with generic three-parameter single-qubit gates, and performed gradient descent in this  multi-dimensional parameter space, to maximize the quantum state fidelity $F_q$ between the output state of the circuit and the four-qubit singlet state.

First, we implement this circuit on IonQ’s trapped ion quantum computer, which has full connectivity and hence can support the Loop Circuit without further compilation. 
These devices have smaller readout and gate error rates compared to IBM Q devices. 
The experiments were run through 
the AWS Braket cloud service \cite{gonzalez2021cloud}. We achieved a classical fidelity $F_c = 0.8923$, and a quantum fidelity of $F_q = 0.885$.
For reference, the classical fidelity between a uniform bitstring distribution and the ideal bitstring distribution is $1 / \sqrt{3} \approx 0.58$.
On the other hand, the quantum fidelity of the ideal four-qubit singlet state and the fully mixed four-qubit state is $\approx 0.0625$.
Hence we conclude that both the classical and quantum fidelities provided by IonQ hardware are much closer to the ideal one than to random noise.
From the experiment on the IonQ hardware, we have inferred a noise strength (as defined in \ref{subsec:robustness}) of $q=0.105$.

\subsection{Linear Circuit}
\begin{figure}[t]
    \centering
    \includegraphics[scale = 0.45]{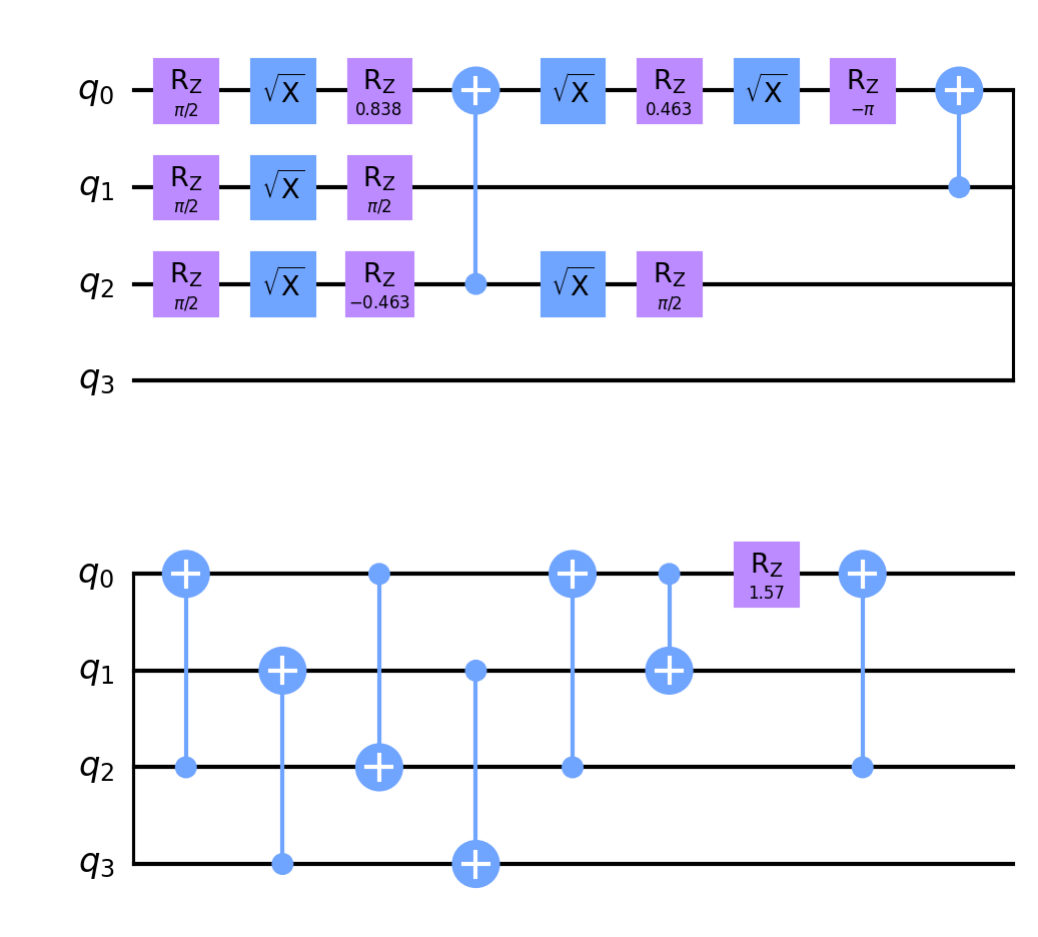}
    \caption{Linear Circuit, a 9-CNOT circuit that
    prepares the four-qubit singlet state
    of Eq.~\eqref{eq:cabello}. The decimal fractions present in the circuit map to 0.83776, 0.46251 and 1.5708.}
    \label{fig:linear_circ}
\end{figure}

The publicly available IBM Q backends (as of August 2023) have four-qubit segments with linear topology, but do not have four-qubit segments with the loop topology. 
Therefore, we have compiled the Loop Circuit to linear topology, to be able to run it on IBM Q backends. 
We call the result the Linear Circuit; it contains 9 CNOT gates, and we show it in  Fig.~\ref{fig:linear_circ}.
We used the Qiskit framework to implement the measurements and carry out a readout error mitigation procedure (see Refs.~\cite{chen2019detector,maciejewski2020mitigation} for theoretical details).
We implemented the Linear Circuit on all available IBM Q devices that support a chain of four qubits in a linear connectivity. The results of the measurements are presented in App.~\ref{app:measurements}.

As for the classical fidelity, our best results were obtained from the qubit chain 3-2-4-1 on IBMQ Manila, yielding $F_c = 0.875$. When we use readout error mitigation, then the best result is found on the qubit chain 1-3-2-5 on IBM Perth, with $F^\mathrm{mitig}_c = 0.996$.

In terms of quantum fidelity, the best performing arrangement is the 3-2-4-1 qubit chain on IBMQ Manila. Without readout error mitigation, the fidelity is $F_q = 0.757$, whereas readout error mitigation improves this to $F_q^{\mathrm{mitig}} = 0.895$. The best quantum fidelity after readout error mitigation is found on the qubit chain 1-3-2-4 on IBMQ Quito, with $F^\mathrm{mitig}_q = 0.931$. As a reference, the quantum fidelity of the ideal four-qubit singlet state and the fully mixed four-qubit state is $\approx 0.0625$.

In terms of the noise strengh $q$ defined in Sec.~\ref{subsec:robustness}, IBMQ Manila had the best performance $1-q = 87.8\%$ without readout error mitigation, implying $q = 12.2\%$. 
Using readout error mitigation, we measured $1-q = 99.96\%$, corresponding to $q = 0.04\%$, on IBM Perth.

To conclude, we identified two circuits that can be used to prepare the four-qubit singlet state. 
We performed measurements on IonQ and IBM Q devices, to quantify the quality of state preparation. 
Our results reveal that the noise level of these devices is low enough that the singlet state can be prepared to a good approximation. 
A more practical question is if a typical resource state prepared by these means has sufficient fidelity to enable a few-percent failure probability of the Weak Broadcast protocol, consuming a few hundred of those resource states?
This is not yet the case in our experiments: taking a typical state preparation error of 1\%, the noisy failure probability (see Eqs.~\eqref{eq:noisy} and \eqref{eq:leakage}) grows above our example target $p_{f,t} = 5$\% already for $m=5$ resource states. 
However, we expect that two orders of magnitude improvement in this state preparation fidelity procedure is required to meet the above target. 

\section{Conclusion and outlook}
\label{sec:discussion}

In this work, we introduced a two-parameter 
family of quantum-aided weak broadcast protocols, 
and proved their security in a certain range of its parameters. 
We computed a tight upper bound $m_{\text{min},\uparrow}$ on the number of resource quantum states required to suppress the failure probability of the protocol below 
a given target failure probability threshold. 
We optimized the protocol, that is, we located the parameter range where the resource requirements, as indicated by $m_{\text{min},\uparrow}$, are minimal.
We have experimentally implemented the preparation
and characterization of our four-qubit
resource state on real qubits: in the superconducting quantum processors of IBM and the ion-trap quantum processors of IonQ. 
Our work illustrates the engineering aspects of future 
deployments of such protocols in practice.
The methods and results of this work can be utilized in 
future research toward multi-party quantum communication in distributed systems.

Let us note that throughout the theoretical part of this work, we assumed that the four-qubit singlet state $\ket{\psi}$ of Eq.~\eqref{eq:cabello} is prepared and distributed through public quantum channels to the three components of the distributed system. The role of these entangled states is to provide random bits that are correlated among the different components. The same functionality cannot be substituted by a source of correlated classical random bits, distributed through public channels. In this case,  the adversary can eavesdrop on the classical channel transmitting the random bits from the source to the components, and hence the criterion that each component knows only its own measurement outcomes can be violated. In the quantum case, considered in this work, the eavesdropping can be detected by sacrificing a fraction of the four-qubit singlet state in a distribute-and-test procedure of the protocol, similar to the procedures described in preceding work \cite{<fitzi_2001>,cabello,cabstateprep}. We leave it for future work to elaborate the quantitative details of such a distribute-and-test procedure.
Naturally, such a distribute-and-test procedure will require extra resources, beyond those estimated in our work.

The above observation points toward further
important follow-up tasks to this work.
For example, an attack against the protocol 
could be based on the adversary gaining control
over the source, or over the measurement
devices generating the correlated random bits.
A further difficulty is posed by physical errors
of the source, the quantum channels, and the measurement
devices, as highlighted by our experiments
presented in Sec.~\ref{sec:experiment}.
We anticipate that the effect of these imperfections
on the protocol will be studied in a framework similar 
to that of device-independent quantum key 
distribution \cite{Acin,Xu_2020}.
Some of these issues have already been 
discussed \cite{fitzi_proceedings,cabello,cabstateprep},
but a quantitative study is, to our knowledge,
still missing. 

\section{Code and Data}
The code and the data we used in this study are available at Zenodo \cite{akosbudai_2023_10364755}. 
Interested readers can access the code and data to replicate and extend the findings of our study.

\acknowledgments

We acknowledge useful discussions with T. Coopmans, 
M. Farkas, M. Fitzi, B. T. Gard, T. Kriv\'achy, and A. Pereszl\'enyi. This research was supported by the Ministry of Innovation and Technology and the National Research, Development and Innovation Office (NKFIH) within the Quantum Information National Laboratory of Hungary, the Quantum Technology National Excellence Program (Project No. 2017-1.2.1-NKP-2017-00001) and the  Thematic Excellence Programmes TKP2020-NKA-06 and TKP2020 IES (Grant No. BME-IE-NAT), under the auspices of the Ministry for Innovation and Technology, and by NKFIH through the OTKA Grants FK 124723, FK 132146, K 124351.
Furthermore, we acknowledge the use of the IBM Q for this work. The views expressed are those of the authors and do not reflect the official policy or position of IBM or the IBM Q team. We acknowledge the access to advanced services provided by the IBM Quantum Researchers Program.

\printbibliography

@article{chen2019detector,
title={{Detector tomography on IBM quantum computers and mitigation of an imperfect measurement}},
author={Chen, Yanzhu and Farahzad, Maziar and Yoo, Shinjae and Wei, Tzu-Chieh},
journal={Physical Review A},
volume={100},
pages={052315},
year={2019},
doi = {10.1103/PhysRevA.100.052315}
}

@article{maciejewski2020mitigation,
title={Mitigation of readout noise in near-term quantum devices by classical post-processing based on detector tomography},
author={Maciejewski, Filip B and Zimbor{\'a}s, Zolt{\'a}n and Oszmaniec, Micha{\l}},
journal={Quantum},
volume={4},
pages={257},
year={2020},
doi={https://doi.org/10.22331/q-2020-04-24-257}
}

@article {Pompili,
title = {Realization of a multinode quantum network of remote solid-state qubits},
author = {Pompili, M. and Hermans, S. L. N. and Baier, S. and Beukers, H. K. C. and Humphreys, P. C. and Schouten, R. N. and Vermeulen, R. F. L. and Tiggelman, M. J. and dos Santos Martins, L. and Dirkse, B. and Wehner, S. and Hanson, R.},
journal = {Science},
volume = {372},
pages = {259--264},
year = {2021},
doi = {10.1126/science.abg1919}
}

@article{Krutyanskiy,
title = {Entanglement of Trapped-Ion Qubits Separated by 230 Meters},
author = {Krutyanskiy, V. and Galli, M. and Krcmarsky, V. and Baier, S. and Fioretto, D. A. and Pu, Y. and Mazloom, A. and Sekatski, P. and Canteri, M. and Teller, M. and Schupp, J. and Bate, J. and Meraner, M. and Sangouard, N. and Lanyon, B. P. and Northup, T. E.},
journal = {Phys. Rev. Lett.},
volume = {130},
pages = {050803},
year = {2023},
doi = {10.1103/PhysRevLett.130.050803}
}

@article{Stephenson,
title = {High-Rate, High-Fidelity Entanglement of Qubits Across an Elementary Quantum Network},
author = {Stephenson, L. J. and Nadlinger, D. P. and Nichol, B. C. and An, S. and Drmota, P. and Ballance, T. G. and Thirumalai, K. and Goodwin, J. F. and Lucas, D. M. and Ballance, C. J.},
journal = {Phys. Rev. Lett.},
volume = {124},
pages = {110501},
year = {2020},
doi = {10.1103/PhysRevLett.124.110501},
}

@article{Storz,
title = {{Loophole-free Bell inequality violation with superconducting circuits}},
author = {Storz, Simon and Sch{\"a}r, Josua and Kulikov, Anatoly and Magnard, Paul and Kurpiers, Philipp and L{\"u}tolf, Janis and Walter, Theo and Copetudo, Adrian and Reuer, Kevin and Akin, Abdulkadir and Besse, Jean-Claude and Gabureac, Mihai and Norris, Graham J. and Rosario, Andr{\'e}s and Martin, Ferran and Martinez, Jos{\'e} and Amaya, Waldimar and Mitchell, Morgan W. and Abellan, Carlos and Bancal, Jean-Daniel and Sangouard, Nicolas and Royer, Baptiste and Blais, Alexandre and Wallraff, Andreas},
journal = {Nature},
volume = {617},
pages = {265--270},
year = {2023},
doi = {https://doi.org/10.1038/s41586-023-05885-0}
}

@article{Magnard,
title = {{Microwave Quantum Link between Superconducting Circuits Housed in Spatially Separated Cryogenic Systems}},
author = {Magnard, P. and Storz, S. and Kurpiers, P. and Sch\"ar, J. and Marxer, F. and L\"utolf, J. and Walter, T. and Besse, J.-C. and Gabureac, M. and Reuer, K. and Akin, A. and Royer, B. and Blais, A. and Wallraff, A.},
journal = {Phys. Rev. Lett.},
volume = {125},
pages = {260502},
year = {2020},
doi = {10.1103/PhysRevLett.125.260502}
}

@article {Wehner,
title = {{Quantum internet: A vision for the road ahead}},
author = {Wehner, Stephanie and Elkouss, David and Hanson, Ronald},
journal = {Science},
volume = {362},
pages = {eaam9288},
year = {2018},
doi = {10.1126/science.aam9288},
}

@article{<lamport>,
title = {{The Byzantine Generals Problem}},
author = {Lamport, Leslie and Shostak, Robert and Pease, Marshall},
journal = {ACM Trans. Program. Lang. Syst.},
volume = {4},
pages = {382–401},
year = {1982},
doi = {10.1145/357172.357176}
}

@article{<pease>,
title = {Reaching Agreement in the Presence of Faults},
author = {Pease, M. and Shostak, R. and Lamport, L.},
journal = {J. ACM},
volume = {27},
pages = {228–234},
year = {1980},
doi = {10.1145/322186.322188},
}

@inproceedings{<benor>,
title = {{Fast Quantum Byzantine Agreement}},
author = {Ben-Or, Michael and Hassidim, Avinatan},
year = {2005},
isbn = {1581139608},
publisher = {Association for Computing Machinery},
address = {New York, NY, USA},
doi = {https://doi.org/10.1145/1060590.1060662},
url = {https://doi.org/10.1145/1060590.1060662},
booktitle = {Proceedings of the Thirty-Seventh Annual ACM Symposium on Theory of Computing},
pages = {481–485},
numpages = {5},
series = {STOC '05}
}

@inproceedings{fitzi_proceedings,
author = {Fitzi, Matthias and Gisin, Nicolas and Maurer, Ueli M. and Rotz, Oliver von},
title = {{Unconditional Byzantine Agreement and Multi-Party Computation Secure against Dishonest Minorities from Scratch}},
year = {2002},
publisher = {Springer-Verlag},
booktitle = {{Proceedings of the International Conference on the Theory and Applications of Cryptographic Techniques: Advances in Cryptology}},
pages = {482–501},
doi = {10.1007/3-540-46035-7_32},
series = {EUROCRYPT '02}
}

@article{<fitzi_2001>,
title = {{Quantum Solution to the Byzantine Agreement Problem}},
author = {Fitzi, Matthias and Gisin, Nicolas and Maurer, Ueli},
journal = {Phys. Rev. Lett.},
volume = {87},
pages = {217901},
year = {2001},
doi = {10.1103/PhysRevLett.87.217901},
}

@phdthesis{fitzi_phd,
author       = {Matthias Fitzi},
title        = {Generalized Communication and Security Models in {B}yzantine Agreement},
year         = 2003,
month        = 3,
note         = {Reprint as vol.~4 of {ETH Series in Information Security and Cryptography}, {ISBN} 3-89649-853-3, {H}artung-{G}orre {V}erlag, {K}onstanz, 2003},
school       = {{ETH Zurich}}
}

@article{cabello,
title = {Solving the liar detection problem using the four-qubit singlet state},
author = {Cabello, Ad\'an},
journal = {Phys. Rev. A},
volume = {68},
pages = {012304},
year = {2003},
doi = {10.1103/PhysRevA.68.012304}
}

@article{cabstateprep,
title = {{Experimental Demonstration of a Quantum Protocol for Byzantine Agreement and Liar Detection}},
author = {Gaertner, Sascha and Bourennane, Mohamed and Kurtsiefer, Christian and Cabello, Ad\'an and Weinfurter, Harald},
journal = {Phys. Rev. Lett.},
volume = {100},
pages = {070504},
year = {2008},
doi = {10.1103/PhysRevLett.100.070504}
}

@article{<pathumsoot>,
title = {Modeling of measurement-based quantum network coding on a superconducting quantum processor},
author = {Pathumsoot, Poramet and Matsuo, Takaaki and Satoh, Takahiko and Hajdu{\v{s}}ek, Michal and Suwanna, Sujin and Van Meter, Rodney},
journal = {Phys. Rev. A},
volume = {101},
pages = {052301},
year = {2020},
doi = {10.1103/PhysRevA.101.052301},
}

@article{<garaymoses>,
title = {{Fully Polynomial Byzantine Agreement for n > 3t Processors in t + 1 Rounds}},
author = {Garay, Juan and Moses, Yoram},
journal = {SIAM Journal on Computing},
volume = {27},
pages = {247-290},
year = {1999},
doi = {10.1137/S0097539794265232}
}

@inproceedings{<pbft>,
title={Practical byzantine fault tolerance},
author={Castro, Miguel and Liskov, Barbara and others},
booktitle={OSDI},
volume={99},
number={1999},
pages={173--186},
year={1999}
}

@inproceedings{<raft>,
title = {{In Search of an Understandable Consensus Algorithm}},
author = {Ongaro, Diego and Ousterhout, John},
year = {2014},
publisher = {USENIX Association},
booktitle = {Proceedings of the 2014 USENIX Conference on USENIX Annual Technical Conference},
pages = {305–320},
numpages = {16},
series = {USENIX ATC'14},
}

@article{<paxos>,
title = {{The Part-Time Parliament}},
author = {Lamport, Leslie},
journal = {ACM Trans. Comput. Syst.},
volume = {16},
pages = {133–169},
year = {1998},
doi = {10.1145/279227.279229},
}

@article{Taherkhani_2017,
title = {{Resource-aware system architecture model for implementation of quantum aided Byzantine agreement on quantum repeater networks}},
author = {Mohammand Amin Taherkhani and Keivan Navi and Rodney Van Meter},
journal = {Quantum Science and Technology},
volume = {3},
pages = {014011},
year = 2017,
doi = {10.1088/2058-9565/aa9bb1},
}

@article{Xu_2020,
title = {Secure quantum key distribution with realistic devices},
author = {Xu, Feihu and Ma, Xiongfeng and Zhang, Qiang and Lo, Hoi-Kwong and Pan, Jian-Wei},
journal = {Rev. Mod. Phys.},
volume = {92},
pages = {025002},
year = {2020},
doi = {10.1103/RevModPhys.92.025002}
}

@article{Acin,
title = {Device-Independent Security of Quantum Cryptography against Collective Attacks},
author = {Ac\'{\i}n, Antonio and Brunner, Nicolas and Gisin, Nicolas and Massar, Serge and Pironio, Stefano and Scarani, Valerio},
journal = {Phys. Rev. Lett.},
volume = {98},
pages = {230501},
year = {2007},
doi = {10.1103/PhysRevLett.98.230501}
}

@misc{ibm,
title = {{IBM Quantum}},
howpublished = {\url{https://quantum-computing.ibm.com/}},
year = {2021}
}

@inproceedings{monroe2021ionq,
title = "{IonQ Quantum Computers: Clear to Scale}",
author = {{Monroe}, Christopher},
booktitle = {{APS March Meeting Abstracts}},
year = 2021,
series = {APS Meeting Abstracts},
volume = {2021},
eid = {P10.002},
pages = {P10.002},
adsurl = {https://ui.adsabs.harvard.edu/abs/2021APS..MARP10002M},
}

@article{preskill2018quantum,
title = {Quantum {C}omputing in the {NISQ} era and beyond},
author = {Preskill, John},
journal = {{Quantum}},
volume = {2},
pages = {79},
year = {2018},
doi = {10.22331/q-2018-08-06-79},
}

@article{hucul2015modular,
title = {Modular entanglement of atomic qubits using photons and phonons},
author = {Hucul, D. and Inlek, I. V. and Vittorini, G. and Crocker, C. and Debnath, S. and Clark, S. M. and Monroe, C.},
journal = {Nature Physics},
volume = {11},
pages = {37--42},
year = {2015},
doi = {10.1038/nphys3150},
}

@article{gonzalez2021cloud,
title = {{Cloud based {QC} with Amazon Braket}},
author = {Gonzalez, Constantin},
journal = {Digitale Welt},
volume = {5},
pages = {14--17},
year = {2021},
doi = {10.1007/s42354-021-0330-z},
}

@software{akosbudai_2023_10364755,
title        = {akosbudai/quantum-byzantine: v0.0.2},
author       = {Akos Budai and Istvan Finta and
          Zoltán Guba},
month        = dec,
year         = 2023,
publisher    = {Zenodo},
doi          = {10.5281/zenodo.10364755},
url          = {https://doi.org/10.5281/zenodo.10364755}
}
\newpage
\onecolumn
\appendix
\section{Broadcast, weak broadcast, and their truth tables}
\label{app:broadcast}

In Sec.~\ref{sec:weakbroadcastprotocol}, we introduced  
the broadcast and the weak broadcast
functionalities of multi-party communication protocols.
Here, we recall their definitions \cite{fitzi_phd}.
Furthermore, we show 
their tabular representations (`truth tables'), 
for the case of $n=3$ components and at most
$t=1$ faulty component.

\begin{mydef}{(broadcast)}
\label{broadcast}
A protocol among \textit{n} components such that one distinct component \textit{S} (the Sender) holds an input value $x_S \in  \{ 0,1 \} $, and all other components (the receivers) eventually decide on an output value in $\{ 0,1 \}$ is said to achieve \textbf{broadcast} if the protocol guarantees the following conditions:
\begin{itemize}[label = {}]
    \item \textit{Validity}: if the Sender is correct then all correct components decide on $y = x_S$.
    \item  \textit{Consistency}: all correct components decide on the same output. 
\end{itemize}

\end{mydef}

The truth table of the broadcast 
functionality is shown in Table \ref{tab:broadcast},
for the special case when the number of components
is $n=3$, and resilience is expected up to $t=1$ 
faulty component.

\begin{table}[t]
\centering
\begin{tabular}{|c|c|c|c|c|c|}
\hline
$S$ & $R_0$ & $R_1$ & no faulty                 & $S$ faulty                  & $R_0$ faulty                 \\ \hline
0 & 0              & 0              & \checkmark & \checkmark & \checkmark \\ \hline
0 & 0              & 1              & $\times$                  & $\times$                  & $\times$                  \\ \hline
0 & 1              & 0              & $\times$                  & $\times$                  & \checkmark \\ \hline
0 & 1              & 1              & $\times$                  & \checkmark & $\times$                  \\ \hline
1 & 0              & 0              & $\times$                  & \checkmark & $\times$                  \\ \hline
1 & 0              & 1              & $\times$                  & $\times$                  & \checkmark \\ \hline
1 & 1              & 0              & $\times$                  & $\times$                  & $\times$                  \\ \hline
1 & 1              & 1              & \checkmark & \checkmark & \checkmark \\ \hline
\end{tabular}
\caption{Truth table of broadcast with 
$n=3$ components, resilient up to $t=1$ faulty component.
Column 1 is the input and output bit of the sender
($S$), column 2 (3) is the output bit of receiver $R_0$
($R_1$).
Columns 4, 5, 6 indicate whether the list
of output bits and the faulty configuration
together satisfies the conditions of 
broadcast or not.
The faulty configuration `$R_1$ faulty' is not shown as it is
analogous to $R_0$ faulty. 
\label{tab:broadcast}}
\end{table}

In the definition of the second functionality, weak broadcast, 
the conditions of broadcast are relaxed to 
some extent, and in addition to 0 and 1, 
a third output value called `abort' ($\perp$) is 
also allowed.
As discussed in the main text, the
weak broadcast functionality is important because
it can be used as a building block to achieve
broadcast \cite{fitzi_phd}.

\begin{mydef}{(weak broadcast)}
\label{def:wb}
A protocol among $n$ components such that one distinct component $S$ (the Sender) holds an input value $x_S \in \{0, 1\} $ and all other components eventually decide on an output value in $\{0,1, \perp \} $ is said to achieve \textbf{weak broadcast} if the protocol guarantees the following conditions: 
\begin{itemize}[label={}]
    \item \textit{Validity}: if the Sender is correct then all correct components decide on $y = x_S$.
    \item  \textit{Consistency}: if any correct component decides on an output $y \in \{0, 1\} $ then all correct components decide on a value in $ \{ y, \perp \}$; that is, either they choose the value $y$ or choose abort.
    \end{itemize}
\end{mydef}

The truth table of the weak broadcast 
functionality is shown in Table \ref{tab:weakbroadcast},
for the special case when the number of components
is $n=3$, and resilience is expected up to $t=1$ 
faulty component.
In the main text, this special case
is denoted by WBC(3,1).
Note that the above definition of weak broadcast does not include the case when both receivers chose abort in the $S$ faulty scenario. However, this scenario should be considered as weak broadcast, because it means that both receivers found that $S$ is a faulty component. Also, note that a faulty $S$ can always make the receivers output abort by sending conflicting messages. Because of this, we consider these scenarios as weak broadcast. Another alternative solution to this would be to modify the \textit{Consistency} condition of the definition of weak broadcast. To do this, one needs to allow for an $y \in \{ 0,1 \perp \} $.

\begin{table}[t]
\centering
\begin{tabular}{|c|c|c|c|c|c|}
\hline
$S$ & $R_0$   & $R_1$   & no faulty                 & $S$ faulty                  & $R_0$ faulty                 \\ \hline
0 & 0       & 0       & \checkmark & \checkmark & \checkmark \\ \hline
0 & 0       & 1       & $\times$                  & $\times$                  & $\times$                  \\ \hline
0 & 0       & $\perp$ & $\times$                  & \checkmark & $\times$                  \\ \hline
0 & 1       & 0       & $\times$                  & $\times$                  & \checkmark \\ \hline
0 & 1       & 1       & $\times$                  & \checkmark & $\times$                  \\ \hline
0 & 1       & $\perp$ & $\times$                  & \checkmark & $\times$                  \\ \hline
0 & $\perp$ & 0       & $\times$                  & \checkmark & \checkmark \\ \hline
0 & $\perp$ & 1       & $\times$                  & \checkmark & $\times$                  \\ \hline
0 & $\perp$ & $\perp$ & $\times$                  & \checkmark & $\times$                  \\ \hline
1 & 0       & 0       & $\times$                  & \checkmark & $\times$                  \\ \hline
1 & 0       & 1       & $\times$                  & $\times$                  & \checkmark \\ \hline
1 & 0       & $\perp$ & $\times$                  & \checkmark & $\times$                  \\ \hline
1 & 1       & 0       & $\times$                  & $\times$                  & $\times$                  \\ \hline
1 & 1       & 1       & \checkmark & \checkmark & \checkmark \\ \hline
1 & 1       & $\perp$ & $\times$                  & \checkmark & $\times$                  \\ \hline
1 & $\perp$ & 0       & $\times$                  & \checkmark & $\times$                  \\ \hline
1 & $\perp$ & 1       & $\times$                  & \checkmark & \checkmark \\ \hline
1 & $\perp$ & $\perp$ & $\times$                  & \checkmark & $\times$                  \\ \hline
\end{tabular}
\caption{Truth table of WBC(3,1), that is, 
weak broadcast with 
$n=3$ components, resilient up to $t=1$ faulty component.
Column 1 is the input and output bit of the sender
($S$), column 2 (3) is the output value of receiver $R_0$
($R_1$).
Columns 4, 5, 6 indicate whether the list
of output values and the faulty configuration
together satisfies the conditions of 
weak broadcast or not.
The faulty configuration `$R_1$ faulty' is not shown as it is
analogous to $R_0$ faulty. 
\label{tab:weakbroadcast}}
\end{table}

\section{Adversary strategies against the Weak Broadcast protocol}
\label{app:adversarystrategies}

In this section, we define adversary strategies, and propose an $S$ faulty strategy and an $R_0$ faulty strategy. 
We prove that the strategies proposed here are optimal, i.e., they maximise the failure probability, in App.~\ref{app:proofsofoptimality}.

As a preliminary for this section, recall that in Sec.~\ref{sec:weakbroadcastprotocol}, we defined 
the Elementary Events as the possible correlated random outcomes
of the measurement of the four-qubit singlet state $\ket{\psi}$ 
of Eq.~\eqref{eq:cabello}.
There are six such outcomes, 0011, 1100, 1010, 0101, 1001, and 0110.
It follows from Eq.~\eqref{eq:cabello} that
the probabilities of 0011 and 1100 are $1/3$, 
whereas the probabilities of the remaining four bitstrings are $1/12$.

Furthermore, we have denoted the number of four-qubit singlet states, available for the weak broadcast of a single bit from $S$, as $m$.
In Sec.~\ref{sec:weakbroadcastprotocol}, we defined 
an Event as the random configuration of the $m$ bitstrings measured on the $m$ distributed four-qubit singlet states by 
the three components. 
Fig.~\ref{fig:sfaultyexamples}a shows an example of an Event for $m=12$
(actually, this is the same Event as in Fig.~\ref{fig:Cabstate_distribution}b).
The first column is the index of the singlet state, and columns
2-4 show the qubit measurement outcomes (Elementary Events) obtained by the three components. 
Note that the number of possible different Events is $6^m$, and for a given $m$, it is straightforward to calculate the occurence probability of each possible Event. For example, the Event in Fig.~\ref{fig:sfaultyexamples} occurs with probability $(1/3)^8 (1/12)^4$.

\subsection{\texorpdfstring{$S$ faulty}{S faulty}}
\label{app:sfaultystrategy}

For concreteness, and without the loss of generality, here we assume that the faulty $S$ is a conscious adversary, which wants to reach failure of Weak Broadcast such that receiver $R_0$ outputs $y_0 = 0$ and receiver $R_1$ outputs $y_1 = 1$.
Hence, the message sent by $S$ to $R_0$ is $x_0 = 0$, whereas the message sent by $S$ to $R_1$ is $x_1 = 1$.

\begin{figure}
    \centering
    \includegraphics[scale = 0.4]{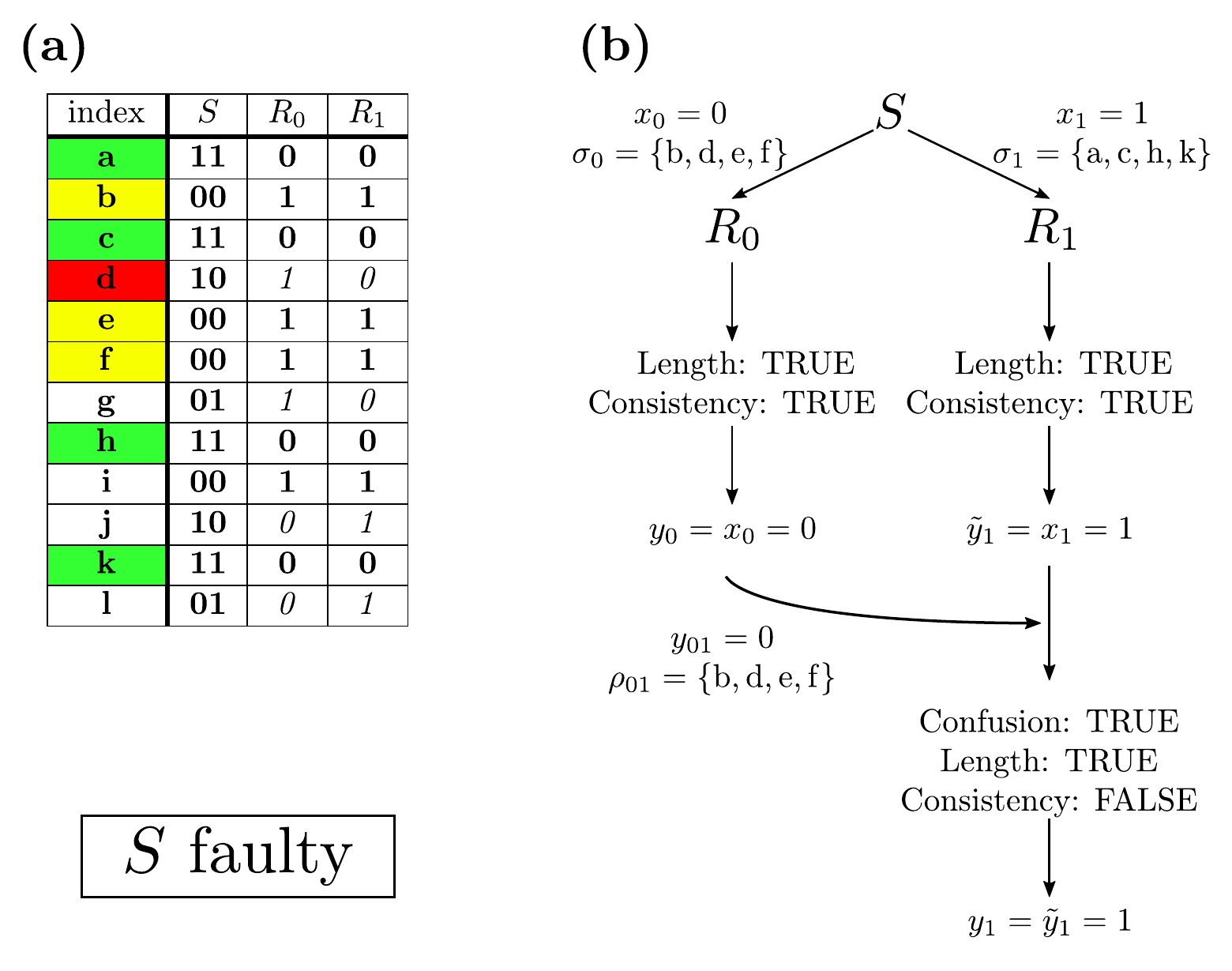}
    \caption{
    An example for a Sender ($S$) 
    adversary strategy, which causes
    failure of Weak Broadcast in this
    particular random instance (Event).
    (a) Random bits resulting from the 
    measurement of $m=12$ four-qubit singlet states.
    Boldface (italic) denotes the information possessed
    by (hidden from) $S$.
    (b) Flowchart of the communication 
    and decision events in case of a faulty $S$
    and correct $R_0$ and $R_1$, for
    the Event in (a) and the 
    adversary strategy
    $\zeta=(3,1,0;0,0,\ell_3)$.
    For this random instance,
    the strategy $\mathcal{S}$ leads
    to failure of Weak Broadcast, as the
    two receivers have different output values, 
    $y_0 \neq y_1$. 
    }
    \label{fig:sfaultyexamples}
\end{figure}

Let us first formalize what we mean by an adversary strategy (of $S$).
To this end, we introduce two natural classifications of the Events; a finer, \emph{global} classification, and a coarser, \emph{local} one. 
Here, the term `local' refers to the property that the classification is based only on the information available to the adversary sender $S$.

We define the \emph{global count list} of an Event as the list of the six non-negative integers that count the frequencies of the six different measurement outcomes of the Event. The set of the global count lists will be denoted as GCL.
That is, an element of GCL is 
\begin{equation}
\label{eq:globalcountlist}
g = (g_1, g_2, g_3, g_4, g_5, g_6) = 
(m_{0011}, m_{0101}, m_{0110}, m_{1001}, m_{1010}, m_{1100}).
\end{equation}
Note that for a certain Event, the adversary $S$ does not know the corresponding global count list $g$, as $S$ knows only the first two bit values of each row of the Event. 

Clearly, for any $g \in \text{GCL}$ it holds that $\sum_{i=1}^6 g_i = m$.
Furthermore, we will denote the counting function that associates an element of GCL to an Event as
\begin{equation}
    G: \text{Events} \rightarrow \text{GCL}.
\end{equation}
Note that this function is not injective, since in general, multiple Events are mapped to a single global count list $g \in \text{GCL}$. 
In fact, for any $g \in \text{GCL}$, the number of Events that are mapped to $g$ via $G$ is 
\begin{equation}
    |G^{-1}(g)| = \binom{m}{g_1,g_2,g_3,g_4,g_5,g_6},
\end{equation}
where the bracket on the right hand side denotes the multinomial coefficient.

An adversary strategy of the sender $S$ describes how $S$ uses its own local information to act during the protocol.
More precisely, the strategy is a procedure, according to which the  adversary $S$ selects the check sets $\sigma_0$ and $\sigma_1$, to be sent to $R_0$ and $R_1$, respectively, in the Invocation Phase.
Locally, the sender $S$ cannot distinguish between the outcomes $0101$ and $0110$, as $S$ knows only the first two bits of these bitstrings. 
Similarly, $S$ cannot distinguish between the outcomes $1010$ and $1001$.
These imply that the basis of an adversary strategy of $S$ is provided by the four non-negative integers $(g_1,g_2+g_3,g_4+g_5,g_6) = 
(m_{0011},m_{0101}+m_{0110},m_{1001}+m_{1010},m_{1100})$.

For our purposes, it is sufficient to use an even coarser local classification, which does not distinguish between the local bit pairs $01$ and $10$ of the sender $S$.
This coarser classification is defined by the map 
\begin{equation}
    \mathcal{L}_S: \text{GCL} \to \text{LCL}_S, \,
    g \mapsto 
    \ell \equiv 
    (\ell_1,\ell_2,\ell_3) \equiv 
    (g_1,g_2+g_3+g_4+g_5,g_6).
\end{equation}
Here, $\text{LCL}_S$ is the set of \emph{local count lists} for the $S$ faulty scenario, which is implicitly defined as the range of $\mathcal{L}_S$, containing triplets of non-negative integers, such that $\ell_1 + \ell_2 + \ell_3 = m$.
The term \emph{local count list} hence refers to the integer triplets $\ell \in \text{LCL}_S$.

\begin{figure}
    \centering
    \includegraphics[scale = 0.5]{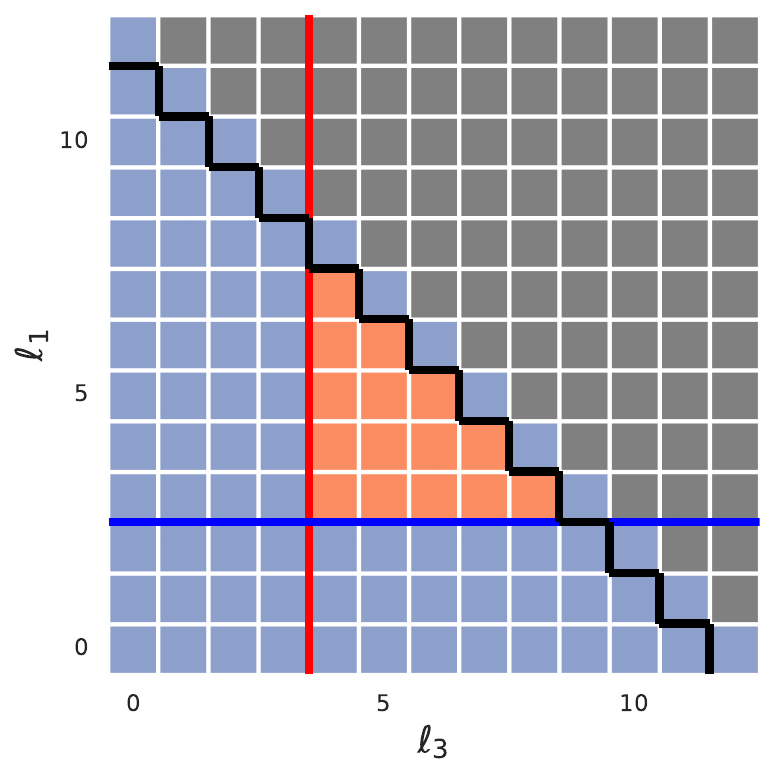}
    \caption{Local count lists and the domain of an incomplete strategy of an adversary Sender ($S$ faulty configuration).
    Each non-grey square depicts a local count list in case of $m=12$.
    Orange region depicts the domain $D(\zeta)$ of the optimal incomplete adversary strategy $\zeta$, for parameter values $\mu = 0.272$, and $\lambda = 0.94$.
    The strategy $\zeta$ is defined in Eq.~\eqref{eq:sstrategy}, whereas the domain $D(\zeta)$ is defined via Eqs.~\eqref{eq:cond1}, \eqref{eq:cond2}, and \eqref{eq:cond3}, depicted as the blue, black, and red lines, respectively.
    Blue region depicts those local count lists where the incomplete strategy $\zeta$ is not defined.}
    \label{fig:sfaulty_strategydomain}
\end{figure}

Because of the sum rule $\ell_1 + \ell_2 + \ell_3 = m$, a local count list $\ell$ is represented by two of its elements, e.g., $\ell_1$ and $\ell_3$, and hence the set $\text{LCL}_S$ can be visualised in a two-dimensional plot. This is illustrated for the case $m=12$ in Fig.~\ref{fig:sfaulty_strategydomain}, where each non-grey square (i.e., orange and blue squares) represents a local count list.

The adversary strategy of $S$ describes how the behavior of $S$ is derived from the local information $S$ has. 
For our purposes, it is sufficent to regard the local count list $\ell$  associated to the Event as the local information possessed by $S$.
We define an adversary strategy of $S$ as a function $\zeta_S$ that maps the local count list of $S$ to a list of 6 non-negative integers, 
\begin{equation}
\label{eq:sstrategydef}
    \zeta_S: \text{LCL}_S \to \left(\mathbb{Z}_0^+\right)^6, \ 
    \ell \mapsto (k^{(0)}_{0011},k^{(0)}_\text{mixed},k^{(0)}_{1100};
    k^{(1)}_{0011},k^{(1)}_\text{mixed},k^{(1)}_{1100}).
\end{equation}
Here, $k^{(0)}_{0011} \leq \ell_1$ indicates how many of the 0011 indices does $S$ include in the check set $\sigma_0$ sent to $R_0$ in the Invocation phase. 
Similarly, $k^{(0)}_\text{mixed} \leq \ell_2$ indicates the number of mixed-outcome (01XX and 10XX) indices to be included in $\sigma_0$, and $k^{(1)}_\text{1100} \leq \ell_3$ indicates the number of 1100 indices to be included in $\sigma_1$, etc.

Our purpose in this work is to analyse 
the failure probability of certain strategies.
For this purpose, it is sufficient to define a strategy by the numbers above: how many
indices from each class are included in each check set.
It is \emph{not} necessary to specify \emph{which} indices are included from each class, because the failure probability is independent of the specific choice.
Nevertheless, in the examples considered in this work, we assume that $S$ includes the `smallest' indices from each class, in alphabetical order of the indices (see Fig.~\ref{fig:sfaultyexamples}). 

We call a strategy, as defined in Eq.~\eqref{eq:sstrategydef}, a \emph{complete strategy}, if its domain is the complete $\text{LCL}_S$; otherwise, we call it an \emph{incomplete strategy}.
For a complete strategy, the failure probability $p_f$ is a well-defined real number $p_f \in [0,1]$. 
From now on, we will deal with incomplete strategies, and hence we will not be able to compute the specific failure probability; however, we can derive a relatively tight upper bound for the failure probability.

In what follows, we will often consider sequences of strategies: such a strategy sequence is a function that specifies a strategy for each $m$.
A natural notation for a strategy in such a sequence $\zeta_{S,m}$; note that we will often suppress the $m$, and call such a sequence of strategies simply as a strategy.
In general, for different strategies in a given sequence, the domain and the failure probability (or the failure-probability upper bound, for incomplete strategies) are different. 
 
In this work, we focus on a sequence of optimal incomplete adversary strategies of $S$. 
This is given as
\begin{equation}
\label{eq:sstrategy}
    \zeta_S(\ell_1, \ell_2, \ell_3) = 
    \left(
        T-Q, Q, 0; 0, 0, \ell_3
    \right),
\end{equation}
where 
\begin{eqnarray}
\label{eq:Tdef}
T&=&\ceil{\mu m},
\\
\label{eq:Qdef}
Q &=& T- \ceil{\lambda T} +1,
\end{eqnarray}
with $\ceil{.}$ denoting the ceiling function.
E.g., for $\mu = 0.26$, $\lambda = 0.94$, and $m=1200$, these are $T=312$ and $Q=19$.
Recall that $T$, originally defined in the Weak Broadcast protocol in Sec.~\ref{sec:weakbroadcastprotocol}, is the minimal check set length that satisfies the Length Conditions of the Weak Broadcast protocol.
Integer $Q$ is the minimal number of inconsistent indices in $\rho_{01}$ that implies the violation of the Consistency Condition of the Cross-check Phase, in the case when the length of the check set is $|\rho_{01}| = T$.

Note that Eq.~\eqref{eq:sstrategy} contains two implicit restrictions on the domain of $\zeta_S$. 
First, the adversary $S$ can include $T-Q$ indices of 0011 outcomes in its check set $\sigma_0$ only if the number $\ell_1$ of 0011 outcomes is sufficient, that is, if
\begin{equation}
\label{eq:cond1}
    T-Q \leq \ell_1.
\end{equation}
Second, the adversary $S$ can include $Q$ indices of mixed outcomes in its check set $\sigma_0$ only if the number $\ell_2$ of mixed outcomes is sufficient, that is, if
\begin{equation}
\label{eq:cond2}
    Q \leq \ell_2.
\end{equation}
Besides these two restrictions, we further restrict the domain of $\zeta$ by requiring 
\begin{equation}
\label{eq:cond3}
    T \leq \ell_3.
\end{equation}
The above 3 conditions define the domain $D(\zeta)$ of the strategy $\zeta$.
With this extra condition, the incomplete strategy $\zeta$ of Eq.~\eqref{eq:sstrategy} is optimal in its domain $D(\zeta)$, which will be proven in App.~\ref{app:sfaultyoptimality}.
The domain $D(\zeta)$ is shown as the orange region in Fig.~\ref{fig:sfaulty_strategydomain}, and the conditions Eqs.~\eqref{eq:cond1}, \eqref{eq:cond2}, and \eqref{eq:cond3} are depicted as the blue, black, and red lines in Fig.~\ref{fig:sfaulty_strategydomain}, respectively.

How can the strategy $\zeta$ in Eq.~\eqref{eq:sstrategy} lead to failure? This is exemplified by a specific instance of the Weak Broadcast protocol in Fig.~\ref{fig:sfaultyexamples}, for $m=12$. 
To be specific, we consider a Weak Broadcast protocol with parameter values $\mu =0.26$ and $\lambda = 0.94$.
The adversary strategy derived from the optimal strategy of Eq.~\eqref{eq:sstrategy} using these values of $\mu$, $\lambda$ and $m$ can be written as 
\begin{equation}
\label{eq:sstrategyexample}
    \zeta = (3,1,0;0,0,\ell_3),
\end{equation}
since $T = 4$ and $Q=1$.

The table in Fig.~\ref{fig:sfaultyexamples}a shows an Event for $m=12$.
The local information of $S$, i.e., the information $S$ possesses, is denoted by boldface characters in Fig.~\ref{fig:sfaultyexamples}a: 
$S$ knows the indices of the four-qubit singlet states
used in the protocol, it knows its own 
measurement outcomes (bit-pairs), and it also knows
the measurement outcomes of $R_0$ and $R_1$ for those rows where
$S$ measured 00 or 11.
The information $S$ does not possess is denoted by italic characters in Fig.~\ref{fig:sfaultyexamples}a.

Based on the local information $S$ has, it classifies the indices $\alpha \in \{1,\dots,m\} \equiv \{\mathrm{a}, \mathrm{b}, \dots, \mathrm{l}\}$ 
of the Event into the three classes defined above: 0011, mixed, 1100.
In the example of Fig.~\ref{fig:sfaultyexamples}a, these
classes can be identified as the corresponding check sets:
$0011 = \{\mathrm{b,e,f,i}\}$, 
$\text{mixed} = \{\mathrm{d,g,j,l}\}$,
and $1100 = \{\mathrm{a,c,h,k}\}$.
This implies $m_{0011} = m_\text{mixed} = m_{1100} = 4$, i.e,
$\ell_1 = \ell_2 = \ell_3 = 4$.
This local count list $\ell = (4,4,4)$ is in the domain, defined by Eqs.~\eqref{eq:cond1}, \eqref{eq:cond2}, \eqref{eq:cond3}, of the incomplete strategy $\zeta$ in Eq.~\eqref{eq:sstrategyexample}.
Hence $S$ can follow the incomplete strategy $\zeta_S$, which results in the communication and decision steps depicted in Fig.~\ref{fig:sfaultyexamples}b.
There, the receivers $R_0$ and $R_1$ follow the Weak Broadcast protocol, and their output bit values are $y_0 = 0$ and $y_1 = 0$, respectively, implying failure of the Weak Broadcast protocol.

The above example sheds light on why the strategy $\zeta_S$ in Eq.~\eqref{eq:sstrategy} is efficient (in fact, optimal in its domain) in achieving failure.
In more general terms, its efficiency is reasoned as follows. 
According to this strategy, the adversary sender $S$ tries to send a check set $\sigma_0$ such that $R_0$ finds it convincingly long (that is, the Length Condition of the Check Phase is satisfied) and
consistent (that is, the Consistency Condition of the Check Phase is satisfied);
but when sent over to $R_1$ as $\rho_{01}$, then $R_1$ finds it inconsistent
(Consistency Condition of the Cross-Check Phase is violated), resulting
in failure via $y_0 = 0$ and $y_1 = 1$.
This is achieved by the strategy $\zeta_S$ of Eq.~\eqref{eq:sstrategy}, if \emph{each} of the $Q$ rows chosen from the mixed class by $S$ contain bit 1 at $R_0$ and bit 0 at $R_1$. 
If at least a single one of these rows contains bit 0 at $R_0$ and bit 1 at $R_1$, then $R_0$ finds the check set $\sigma_0$ inconsistent with the data bit $x_0 = 0$ and hence produces an output $y_0 = \perp$, leading to Weak Broadcast (see Table \ref{tab:weakbroadcast}, 
row 8 or row 17). 

\subsection{\texorpdfstring{$R_0$ faulty}{R0 faulty}}
\label{app:rfaultystrategy}

For concreteness, and without the loss of generality, here we assume that the correct sender $S$ sends the message $x_S = x_0 = x_1 = 0$ to the receivers. 
Furthermore, we assume that the faulty $R_0$ is a conscious adversary, which wants to reach failure of Weak Broadcast such that the correct $R_1$ has an output bit value $y_1 =1$ that conflicts with the  output bit value of $x_S = 0$ of $S$.

We start by formalizing what we mean by an adversary strategy of $R_0$.
Similarly to the $S$ faulty configuration, our starting point is the set GCL of global count lists, see Eq.~\eqref{eq:globalcountlist}.
The adversary strategy of the receiver $R_0$ describes how $R_0$ uses its own local information to act during the protocol. 
Then, locally, the receiver $R_0$ can distinguish between three types of outcomes. 
First, if the third bit of the outcome, i.e., the bit $R_0$ has measured, carries the value `1', and the index of the outcome is sent from $S$ to $R_0$ in the check set $\sigma_0$, then   $R_0$ can identify the outcome as 0011.
Second, if the third bit of the outcome carries the value `1', and the index of the outcome is not included in $\sigma_0$, then $R_0$ identifies the outcome as `either 0110 or 1010'. 
In any other case, i.e., if the third bit of the outcome carries the value of `0', then $R_0$ identifies the outcome as `either 0101, or 1001, or 1100'.
It is natural to denote the frequency of each type of outcome in a given Events as $m_{0011}$, $m_\text{XX10}$, and $m_\text{XX0X}$.

This threefold classification of the Events according to $R_0$'s local information is therefore defined by the map
\begin{equation}
    \mathcal{L}_R\, : \, 
    \text{GCL} \to \text{LCL}_R,~
    g \mapsto \ell \equiv (\ell_1,\ell_2,\ell_3)
    \equiv
    (m_{0011}, m_\text{XX10}, m_\text{XX0X})
    \equiv 
    (g_1,g_3+g_5,g_2+g_4 + g_6).
\end{equation}
Here, $\text{LCL}_R$ is the set of local count lists for the $R_0$ faulty scenario, which is implicitly defined as the range of $\mathcal{L}_R$, containing triplets of non-negative integers, such that $\ell_1 + \ell_2 + \ell_3 = m$.
The term \emph{local count list} hence refers to the integer triplets $\ell \in \text{LCL}_R$.

\begin{figure}
    \centering
    \includegraphics[scale = 0.5]{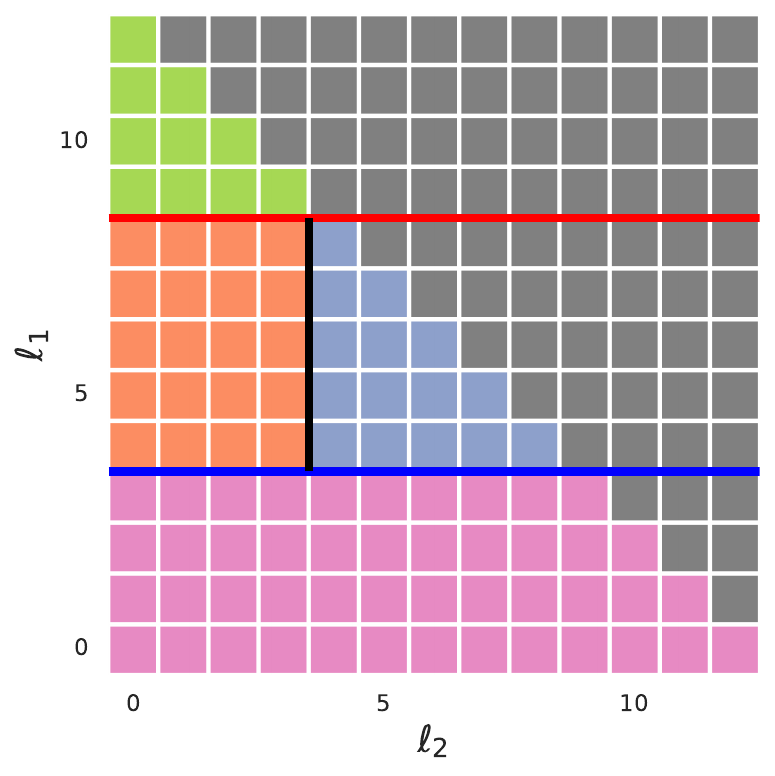}
    \caption{Local count lists and the domain of an incomplete optimal strategy of an adversary receiver $R_0$ 
    ($R_0$ faulty configuration).
    Each non-grey square depicts a local count list in the case of $m=12$. 
    The union of orange, blue and pink regions depicts the domain $D(\zeta)$ of the optimal incomplete strategy $\zeta$, for parameter values $\mu = 0.272$ and $\lambda = 0.94$.
    The strategy $\zeta$ is defined in Eq.~\eqref{eq:r0faultydef}, whereas the domain is defined by Eq.~\eqref{eq:r0faultydomain}.
    Green region depicts local count list where the incomplete strategy $\zeta$ is not defined.}
    \label{fig:r0faulty_stratdom}
\end{figure}

Because of the sum rule $\ell_1 + \ell_2 + \ell_3  = m$, a local count list $\ell$ is represented by two of its elements, e.g., $\ell_1$ and $\ell_2$, and hence the set $\text{LCL}_R$ can be visualised in a two-dimensional plot. 
This is illustrated for the case $m=12$ in Fig.~\ref{fig:r0faulty_stratdom}, where each non-grey square represents a local count list. 

The adversary strategy of $R_0$ describes the procedure to assemble the check set $\rho_{01}$, based on the local information $R_0$ has. 
Hence, we define an adversary strategy of $R_0$ as a function $\zeta_R$ that maps the local count list of $R_0$ to a list of 3 non-negative integers:
\begin{equation}
\label{eq:r0strategy}
    \zeta_R: \text{LCL}_R \to \left(\mathbb{Z}_0^+\right)^3, 
    \ell \mapsto \left(
        k_{0011},k_\text{XX10}, k_\text{XX0X}
    \right).
\end{equation}
Here, $k_{0011} \leq \ell_1$ indicates how many of the 0011 indices does $R_0$ include in the check set $\rho_{01}$ sent to $R_1$ in the Cross-calling Phase. 
Similarly, $k_\text{XX10} \leq \ell_2$ indicates the number of XX10 indices to be included in $\rho_{01}$, and $k_\text{XX0X} \leq \ell_3$ indicates the number of XX0X indices to be included in $\rho_{01}$.

\begin{figure}
    \centering
    \includegraphics[scale = 0.4]{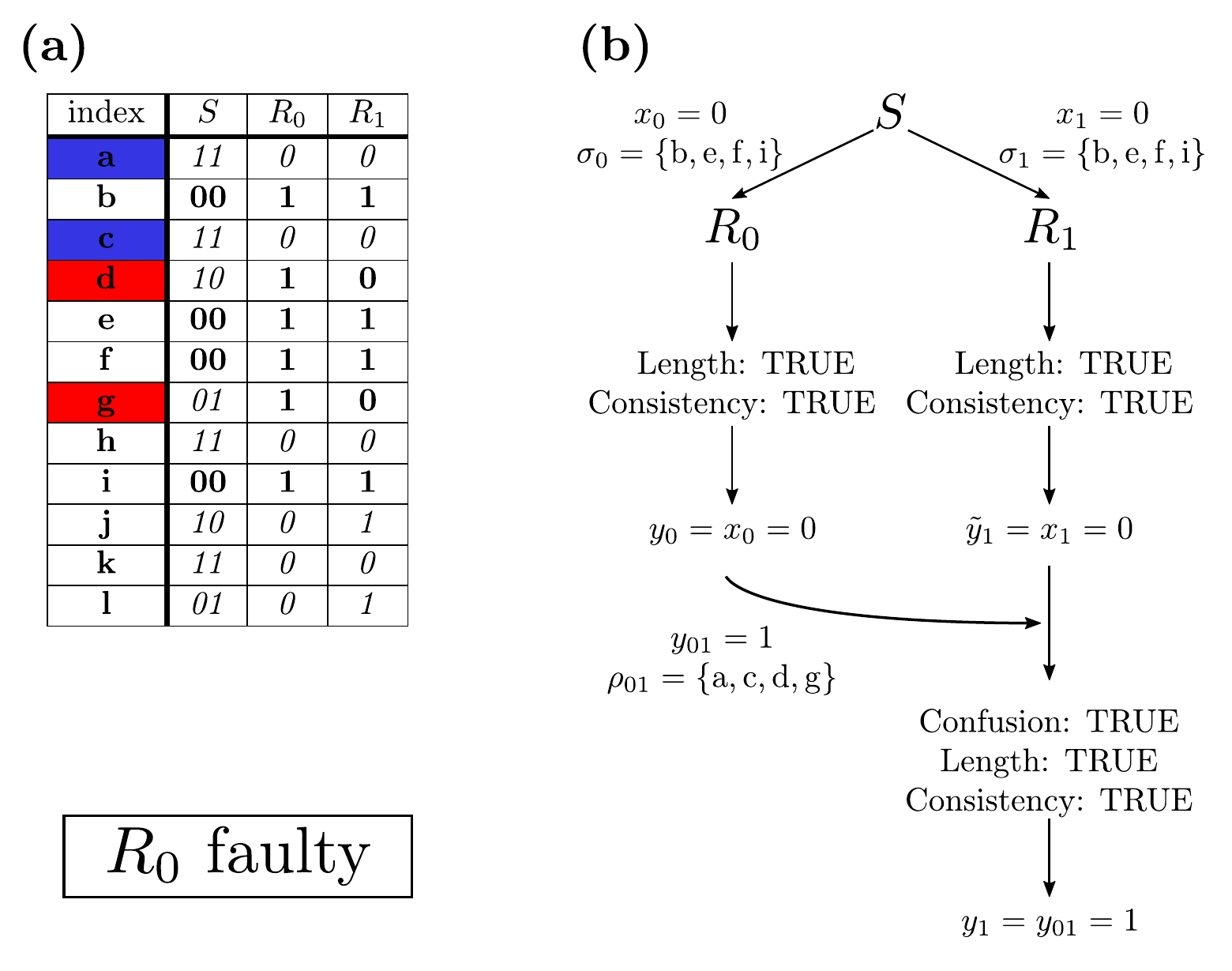}
    \caption{An example for a Receiver ($R_0$) adversary strategy,
    which causes failure of Weak Broadcast in this particular random instance (Event). 
    (a) Random bits resulting from the measurement of $m=12$ four-qubit
    singlet states.
    Boldface (italic) denotes the information possessed by (hidden from)
    $R_0$. 
    (b) Flowchart of the communication and decision steps in case of a 
    faulty $R_0$ and correct $S$ and $R_1$, for the Event in (a) and 
    the adversary strategy $\mathcal{S} = (0,\ell_2,n_\text{min})$, see Eq.~\eqref{eq:r0faultydef}. 
    For this random Event, the strategy $\zeta_R$ leads to 
    failure of Weak Broadcast, as the two receivers have different
    outputs values, $y_0 \neq y_1$.
    }
    \label{fig:r0faultyexamples}
\end{figure}

Similarly to the case of the $S$ adversary strategy discussed above, here it is also sufficient to define a strategy by the `$k$' numbers above in Eq.~\eqref{eq:r0strategy}: 
how many indices from each class are included in each check set.
It is \emph{not} necessary to specify \emph{which} indices are included from each class, because the failure probability is independent of the specific choice.
Nevertheless, in the examples considered in this work, we assume that $R_0$ includes the `smallest' indices from each class, in alphabetical order of the indices (see Fig.~\ref{fig:r0faultyexamples}). 

We call a strategy, as defined in Eq.~\eqref{eq:r0strategy}, a complete strategy, if its domain is the complete $\text{LCL}_R$; otherwise, we call it an incomplete strategy. 
From now on, we deal with incomplete strategies, and hence we will not be able to compute the specific failure probability; however, we will derive an upper bound for the failure probability. 

We consider the incomplete strategy defined as
\begin{equation}
\label{eq:r0faultydef}
\zeta_R(\ell_1,\ell_2,\ell_3) = (0,\ell_2,n_\text{min}),
\end{equation}
where $n_\text{min}$ is the smallest non-negative integer such that $\ell_2+n_\text{min} \geq T$. 
That is,
\begin{equation}
\label{eq:nmindef}
n_\text{min} = 
\begin{cases}
T-\ell_2 & \text{if}~ \ell_2 < T \\
0 & \mbox{if }\ell_2 \geq T.
\end{cases}
\end{equation}

The above definition of $\zeta_R$ contains an implicit assumption on the domain of $\zeta_R$:
the strategy can be followed only if the value of $n_\text{min}$, as defined in Eq.~\eqref{eq:nmindef}, is not greater than $\ell_3$.
More precisely, if $\ell_2 < T$, then $T-\ell_2 \leq \ell_3$ must hold.
Formulating this in terms of $\ell_1$ and $\ell_2$: 
if $\ell_2 < T$, then $\ell_1 \leq m-T$ must hold.
As $\ell_2 \geq T$ implies $\ell_1 \leq m-T$, we conclude that the domain of the above strategy is defined by
\begin{equation}
\label{eq:r0faultydomain}
    \ell_1 \leq m-T.
\end{equation}
In Fig.~\ref{fig:r0faulty_stratdom}, this domain $D(\zeta_R)$ is illustrated for a specific choice of parameters (see caption). 
There, the domain $D(\zeta_R)$ of the strategy is shown as the union of the orange, blue and pink regions, whereas the green region is the complement $\bar{D}(\zeta_R)$ of the domain.
Furthermore, the condition \eqref{eq:r0faultydomain} is depicted as the red line.

We claim that the incomplete strategy $\zeta_R$ is optimal on its domain. 
This will be proven in App.~\ref{app:r0faultyoptimality}.

How can the strategy $\zeta_R$ of Eq.~\eqref{eq:r0faultydef} lead to failure? This is exemplified by a specific instance of the Weak Broadcast protocol in Fig.~\ref{fig:r0faultyexamples}, for $m=12$. 
To be specific, we consider a Weak Broadcast protocol with parameter values $\mu =0.26$ and $\lambda = 0.94$.
The table in Fig.~\ref{fig:r0faultyexamples}a shows an Event for $m=12$.
The local information possessed by $R_0$ is denoted by boldface characters in Fig.~\ref{fig:r0faultyexamples}a (cf.~the second paragraph of this section); the information hidden from $R_0$ is denoted as italic.

Based on the local information $R_0$ has, it classifies the indices $\alpha \in \{1,\dots,m\} \equiv \{\mathrm{a}, \mathrm{b}, \dots, \mathrm{l}\}$ 
of the Event into the three classes defined above: 0011, XX10, XX0X.
In the example of Fig.~\ref{fig:r0faultyexamples}a, these
classes can be identified as the corresponding check sets:
$0011 = \{\mathrm{b,e,f,i}\}$, 
$\text{XX10} = \{\mathrm{d,g}\}$,
and $\text{XX0X} = \{\mathrm{a,c,h,j,k,l}\}$.
This implies $m_{0011} \equiv \ell_1 = 4$, 
$m_\text{XX10} \equiv \ell_2 = 2$, and $m_\text{XX0X} \equiv \ell_3 = 6$.
This local count list $\ell = (4,2,6)$ is in the domain $D(\zeta_R)$ of the incomplete strategy $\zeta_R$; in fact, it is in the orange region of Fig.~\ref{fig:r0faulty_stratdom}.
Hence $R_0$ can follow the incomplete strategy $\zeta_R$, which results in the communication and decision steps depicted in Fig.~\ref{fig:r0faultyexamples}b.
There, the sender $S$ and the receiver $R_1$ follow the Weak Broadcast protocol, and their output bit values are $x_S = 0$ and $y_1 = 1$, respectively, implying failure of the Weak Broadcast protocol.

The above example sheds light on why the strategy $\zeta_R$ in Eq.~\eqref{eq:r0strategy} is efficient (in fact, optimal in its domain) in achieving failure.
In more general terms, its efficiency is reasoned as follows. 
According to this strategy, the adversary receiver $R_0$ sends a false data bit value ($y_{01} = 1 \neq 0 = x_S$) and tries to `back it up' by a check set $\rho_{01}$ that is convincingly long (that is, the Length Condition of the Cross-Check Phase is satisfied) and consistent (that is, to achieve that the Consistency Condition of the Cross-Check Phase is satisfied).
This is achieved by the strategy $\zeta_R$ of Eq.~\eqref{eq:r0strategy}, if the number of consistent indices is large enough; hence it is reasonable for $R_0$ to include \emph{all} indices that are certainly consistent with the bit value $y_{01}=1$ (red indices in Fig.~\ref{fig:r0faultyexamples}), and include the minimal amount of indices which are only potentially consistent with the bit value $y_{01} = 1$ (these are the two blue indices in Fig.~\ref{fig:r0faultyexamples}b), such that the Length Condition is satisfied.

\section{Failure probabilities}
\label{app:failureprobabilities}

In this section, we compute the exact failure probability for the `no faulty' adversary configuration, and compute tight upper bounds of the failure probabilities for the adversary configurations `$S$ faulty' and `$R_0$ faulty'. 
These results allow future users of the protocol to allocate resources sparingly. 
In fact, these results are analysed, and are used to optimize the protocol, in the main text.

\subsection{No faulty}

In this subsection, we consider the no faulty configuration. 
Without the loss of generality, assume that the data bit sent by $S$ is $x_S = 0$.
Then, we have the following exact result for the failure probability $p_f$, which we denote as $p_f^{(\textrm{nf})}$:
\begin{equation}
\label{eq:pfnofaulty}
    p_f^{(\mathrm{nf})} 
    = \sum_{m_{0011}=0}^{T-1} \binom{m}{m_{0011}} \left( \frac{1}{3} \right) ^{m_{0011}} \left( \frac{2}{3} \right)^{m-m_{0011}}.
\end{equation}
Recall the definition $T = \ceil{\mu m}$, introduced in the Weak Broadcast protocol, Sec.~\ref{sec:weakbroadcastprotocol}.
Recall also that $T$ is the minimal check set length that satisfies the Length Conditions. 

To derive this result, it is sufficient to note that the protocol leads to failure only if the Length Condition of the Check Phase evaluates as False, that is, if the number of 0011 outcomes is less than the minimal check set length that satisfies the Length Conditions; i.e. if the following condition holds:
\begin{equation}
\label{eq:nofaultycondition}
    m_{0011} < T. 
\end{equation}
Since the Elementary Event 0011 is generated with a probability of $1/3$, the probability of
Eq.~\eqref{eq:nofaultycondition} holding true can be expressed by the binomial formula Eq.~\eqref{eq:pfnofaulty}. 
This concludes the proof.
Note that the analysis generalises straightforwardly to the case when the message of the sender $S$ is $x_S = 1$, and yields the same result for $p_f^{(nf)}$.

\subsection{\texorpdfstring{$S$ faulty}{S faulty}}
\label{app:sfaultyfailureprobability}

Here, we provide an upper bound $p^{(S)}_{f, \uparrow}$ for the failure probability in the $S$ faulty configuration. 
To express this upper bound, 
recall that $T = \ceil{\mu m}$
is the minimal check set length that satisfies the Length Conditions, which was introduced in Eq.~\eqref{eq:Tdef}.
Also recall that the integer $Q$ was defined in Eq.~\eqref{eq:Qdef}. Here we use its exact definition, that is,
$Q = T- \ceil{T \lambda }+1$.
With these notations, the upper bound $p^{(S)}_{f, \uparrow}$ reads
\begin{eqnarray}
\label{eq:ps_upper}
p_{f,\uparrow}^{(S)} &=&  p_{f,\downarrow}^{(S)}+
\left( 1- \sum_{\ell_3=T}^{m-T} \; \sum_{\ell_1=T-Q}^{m-Q-\ell_3}  \binom{m}{\ell_3,\ell_1,m-\ell_1-\ell_3} \left (\frac{1}{3} \right )^{m} \right),
\end{eqnarray}
with 
\begin{eqnarray}
\label{equation:ps_lower}
p_{f,\downarrow}^{(S)} &=& \sum_{\ell_3=T}^{m-T} \; \sum_{\ell_1=T-Q}^{m-Q-\ell_3}  \binom{m}{\ell_3,\ell_1,m-\ell_1-\ell_3} \left (\frac{1}{3} \right )^{m}  2^{-Q}.
\end{eqnarray}

In the derivation of this upper bound, first we express the upper bound formally for a general incomplete strategy, and then we apply it to the $S$ faulty strategy defined above. 

Assume that we have identified an incomplete strategy $\zeta$ that is optimal in its domain $D(\zeta) \subset \text{LCL}_S$. 
An upper bound of the failure probability of this strategy can be expressed as 
\begin{equation}
\label{eq:psuppersimple}
    p^{(S)}_{f,\uparrow} = 
    \sum_{\ell \in D(\zeta)} 
    P_\zeta(\text{failure} | \ell)
    P(\ell) +
    \sum_{\ell \in \bar{D}(\zeta)} 
    P(\ell).
\end{equation}
Here, $P(\ell)$ is the probability that a random Event produces the local count list $\ell$.
Furthermore, $P_\zeta(\text{failure}| \ell)$ is the conditional probability of failure, given the random Event produces the local count list $\ell$. 
Finally, $\bar{D}(\zeta)$ is the complement of $D(\zeta)$, that is, $\bar{D}(\zeta) = \text{LCL}_S \setminus D(\zeta)$.

In words, the formula \eqref{eq:psuppersimple} for the upper bound  is interpreted as follows: the random Event either produces a local count list $\ell$ that is in the domain of the strategy $\zeta$, and then we compute the corresponding failure probability exactly (first sum); or, the Event produces a local count list $\ell$ that is outside the domain of the strategy $\zeta$, and then we assume a worst-case scenario, i.e., that failure happens with unit probability (second sum). 

Using the normalization condition
\begin{equation}
1 = \sum_{\ell \in \text{LCL}_S}  
P(\ell)
= \sum_{\ell \in D(\zeta)} P(\ell)
+ \sum_{\ell \in \bar{D}(\zeta)}  P(\ell),
\end{equation}
we see that Eq.~\eqref{eq:psuppersimple} can be reformulated as
\begin{equation}
\label{eq:puppergeneral2}
p^{(S)}_{f,\uparrow} = 
    \sum_{\ell \in D(\zeta)} 
    P_\zeta(\text{failure} | \ell)
    P(\ell) +
    \left(
    1- 
    \sum_{\ell \in D(\zeta)} 
    P(\ell)
    \right).
\end{equation}

Note that these considerations suggest that for a given incomplete strategy, a lower bound $p^{(S)}_{f,\downarrow}$ of failure probability can also be expressed, by assuming a best-case scenario, i.e., zero failure probability, for Events that produce local count lists that are outside of the domain of the strategy. 
Formally, this lower bound is expressed as
\begin{equation}
\label{eq:pslowerwithell}
p^{(S)}_{f,\downarrow} = 
\sum_{\ell \in D(\zeta)} 
P_\zeta(\text{failure} | \ell)
P(\ell).
\end{equation}

To evaluate the upper bound $p^{(S)}_{f,\uparrow}$ of the failure probability for this optimal partial strategy $\zeta$ via Eq.~\eqref{eq:puppergeneral2}, we first express
\begin{equation}
\label{eq:Pell}
    P(\ell) =
    \binom{m}{\ell_1, m-\ell_1-\ell_3, \ell_3} \left (\frac{1}{3} \right )^{m}.
\end{equation}
Recall that $\ell = (\ell_1,\ell_2,\ell_3)$ is determined by two of its components, e.g., $\ell_1$ and $\ell_3$, because $\ell_1 + \ell_2 + \ell_3 = m$.
Equation \eqref{eq:Pell} is a straighforward consequence of the fact that the probability of a 0011 outcome, the probability of a mixed outcome (0101 or 0110 or 1001 or 1010), and the probability of a 1100 outcome, are all equal to $1/3$. 
Finally, we express the failure probability of $\zeta$ on its domain, which is 
\begin{equation}
\label{eq:conditional}
    P_\zeta(\text{failure} | \ell) = (1/2)^Q, \mbox{ for all $\ell \in D(\zeta)$}.
\end{equation}
This follows from the fact that $S$ includes $Q$ mixed indices in its check set $\sigma_0$, and this leads to failure only if all the $Q$ corresponding mixed outcomes contain a `1' at $R_0$.
Inserting Eqs.~\eqref{eq:Pell} and \eqref{eq:conditional} into Eq.~\eqref{eq:puppergeneral2}, we obtain the result Eq.~\eqref{equation:ps_lower}, which concludes our derivation.
Note also that the first term of Eq.~\eqref{equation:ps_lower} is in fact the failure-probability lower bound of our specific strategy $\zeta$, in line with the general expression in Eq.~\eqref{eq:pslowerwithell}.

\subsection{\texorpdfstring{$R_0$ faulty}{R0 faulty}}
\label{app:rfaultyfailureprobability}

Here, we provide an upper bound $p^{(R)}_{f,\uparrow}$ for the failure probability in the $R_0$ faulty configuration. 
Again, recall that $T = \ceil{\mu m}$ (original definition in Eq.~\eqref{eq:Tdef}) and 
$Q = T- \ceil{T \lambda }+1$ (original definition in Eq.~\eqref{eq:Qdef}. 
With these notations, the failure-probability upper bound reads:
\begin{equation}
\label{equation:pr0_upper}
\begin{aligned}
p_{f,\uparrow}^{(R)} = p_{f,\downarrow}^{(R)} + \sum_{\ell_1 = m- T+1}^{m} \binom{m}{\ell_1} \left (\frac{1}{3} \right )^{\ell_1}  \left (\frac{2}{3} \right )^{m-\ell_1},
\end{aligned}
\end{equation}
where
\begin{equation}
\label{equation:pr0_lower}
\begin{aligned}
p_{f,\downarrow}^{(R)}  &=\sum_{\ell_1=T}^{m-T} \sum_{\ell_2=0}^{T-Q} \binom{m}{\ell_1,\ell_2,\ell_3} \left(\frac{1}{3} \right)^{\ell_1}  \left(\frac{1}{6} \right)^{\ell_2}  \left(\frac{1}{2}  \right)^{\ell_3} \sum_{k = T- Q + 1 - \ell_2}^{T-\ell_2} \binom{T-\ell_2}{k} \left(\frac{2}{3} \right)^k \left(\frac{1}{3} \right)^{T-\ell_2-k} + \\
&+ \sum_{\ell_1=T}^{m-T} \sum_{\ell_2=T-Q+1}^{m-\ell_1} \binom{m}{\ell_1,\ell_2,\ell_3} \left(\frac{1}{3} \right)^{\ell_1}  \left(\frac{1}{6} \right)^{\ell_2}  \left(\frac{1}{2}  \right)^{\ell_3}+\\
&+ \sum_{\ell_1=0}^{T-1} \binom{m}{\ell_1} \left( \frac{1}{3} \right) ^{\ell_1} \left( \frac{2}{3} \right)^{m-\ell_1}.
\end{aligned}
\end{equation}
The result \eqref{equation:pr0_upper} is valid both for the $x_S = 0$ and the $x_S=1$ scenario. 
In what follows, we derive this result. The derivation follows the 3-step scheme outlined in the $S$ faulty case below Eq.~\eqref{equation:ps_lower}.

Assume that we have identified an incomplete strategy $\zeta$ that is optimal on its domain $D(\zeta) \subset \text{LCL}_R$. 
An upper bound of the failure probability of this strategy can be expressed as shown in Eqs.~\eqref{eq:psuppersimple}.
A lower bound can also be expressed as Eq.~\eqref{eq:pslowerwithell}.

For our optimal incomplete strategy $\zeta$, the formula for the failure-probability upper bound, Eq.~\eqref{eq:psuppersimple}, can be structured further, according to the three regions (orange, pink, blue) in Fig.~\ref{fig:r0faulty_stratdom}:
\begin{equation}
    p^{(R)}_{f,\uparrow} = 
    \sum_{\ell \in \text{orange}}
    P_\zeta(\text{failure}|\ell) P(\ell)
    +
    \sum_{\ell \in \text{pink}}
    P_\zeta(\text{failure}|\ell) P(\ell)
    +
    \sum_{\ell \in \text{blue}}
    P_\zeta(\text{failure}|\ell) P(\ell)
    +
    \sum_{\ell \in \bar{D}(\ell)}
    P(\ell).
\end{equation}
A simplification is allowed by the observation that for local count lists ($\ell$) in the pink and blue regions, the failure probability is $P_\ell(\text{failure}|\ell) = 1$, 
hence
\begin{equation}
\label{eq:pfcolorsimplified}
    p^{(R)}_{f,\uparrow} = 
    \sum_{\ell \in \text{orange}}
    P_\zeta(\text{failure}|\ell) P(\ell)
    +
    \sum_{\ell \in \text{pink}}
    P(\ell)
    +
    \sum_{\ell \in \text{blue}}
    P(\ell)
    +
    \sum_{\ell \in \bar{D}(\ell)}
    P(\ell).
\end{equation}
In the pink region, failure is guaranteed, in fact, for any $R_0$ adversary strategy, since $\ell_1 < T$ implies that the check list $\sigma_0 = \sigma_1$ sent by $S$ violates the Length Condition of the Check Phase. 
In the blue region, however, failure is guaranteed because the adversary $R_0$ was lucky to find enough outcomes XX10, that is, $\ell_2 \geq T-Q+1$, such that its check set $\rho_{01}$ automatically satisfies the Consistency Condition of the Cross-check Phase. 

The last three terms of Eq.~\eqref{eq:pfcolorsimplified} can be readily expressed by binomial and trinomial distributions.
The blue, pink, and green regions correspond to, respectively, to the second term of Eq.~\eqref{equation:pr0_lower}, to the third term of Eq.~\eqref{equation:pr0_lower}, and to the second term of Eq.~\eqref{equation:pr0_upper}.

To complete the derivation of the failure-probability upper bound, we express $P_\zeta(\text{failure}| \ell)$ for $\ell \in \text{orange}$:
\begin{equation}
\label{eq:orangefailure}
    P_\zeta(\text{failure} | \ell) = \sum_{k = T-Q+1-\ell_2}^{T-\ell_2} \binom{T-\ell_2}{k} \left(\frac{2}{3} \right)^k \left(\frac{1}{3} \right)^{T-\ell_2-k}, \mbox{ for $\ell \in$ orange}.
\end{equation}
To interpret this formula, let us first recall that in the orange region of $D(\zeta)$, it holds that $n_\text{min} = T-\ell_2$. 
Furthermore, note also that the probability that the number of XX10 outcomes in the check set $\rho_{01}$ of length $n_\text{min} = T- \ell_2$ is $k$ is given by the binomial distribution.
Hence, Eq.~\eqref{eq:orangefailure} sums up the probabilities where $k$ is large enough that $\rho_{01}$ satisfies the Consistency Condition of the Cross-check Phase. 
Combining Eq.~\eqref{eq:orangefailure} with the trinomial distribution expressing the probability of the local count lists of the orange region, we obtain the first term of Eq.~\eqref{equation:pr0_lower}. This concludes the derivation of the failure-probability upper bound in the case of an adversary $R_0$.

\section{Proofs of optimality}
\label{app:proofsofoptimality}

\subsection{\texorpdfstring{$S$ faulty}{S faulty}}
\label{app:sfaultyoptimality}

Here, we show that the $S$ faulty adversary strategy defined above is an optimal incomplete strategy, i.e., it is optimal on its domain. 

Consider therefore an arbitrary strategy
\begin{eqnarray}
\mathcal{\zeta}_B = \left(
k^{(0)}_{0011},
k^{(0)}_\text{mixed},
k^{(0)}_{1100};
k^{(1)}_{0011},
k^{(1)}_\text{mixed},
k^{(1)}_{1100}
\right).
\end{eqnarray}
It is straightforward to see that the strategy
\begin{eqnarray}
\mathcal{\zeta}'_B = \left(
k^{(0)}_{0011},
k^{(0)}_\text{mixed},
0;
0,
0,
m_{1100}
\right),
\end{eqnarray}
derived from $\mathcal{S}_B$,
has a higher or equal failure probability than the initial Bad Strategy $\mathcal{S}_B$:
$p_f(\mathcal{S}_B) \leq p_f(\mathcal{S}'_B)$.
The reason is as follows.
On the one hand, deviating from 
$\left(
k^{(1)}_{0011},
k^{(1)}_\text{mixed},
k^{(1)}_{1100}
\right) = (0,0,m_{1100})$
either decreases the failure probability, or does not change that. 
In words: from the viewpoint of an adversary $S$, 
it is useless to deviate
from the correct communication
toward $R_1$.
On the other hand, $k^{(0)}_{1100} > 0$ leads to weak broadcast, implying $p_f = 0$.

To analyze the failure probability of
the `improved' strategy $\mathcal{\zeta}'_B$ 
further, we introduce
$Q' = k^{(0)}_\text{mixed}$ and
$T' = k^{(0)}_{0011} + Q'$.
We address the question:
are any strategies defined by $(T',Q') \neq (T,Q)$ better than the
strategy the strategy $\zeta$?
No, as we argue now. 

\begin{figure}
    \centering
    \includegraphics[scale = 0.25]{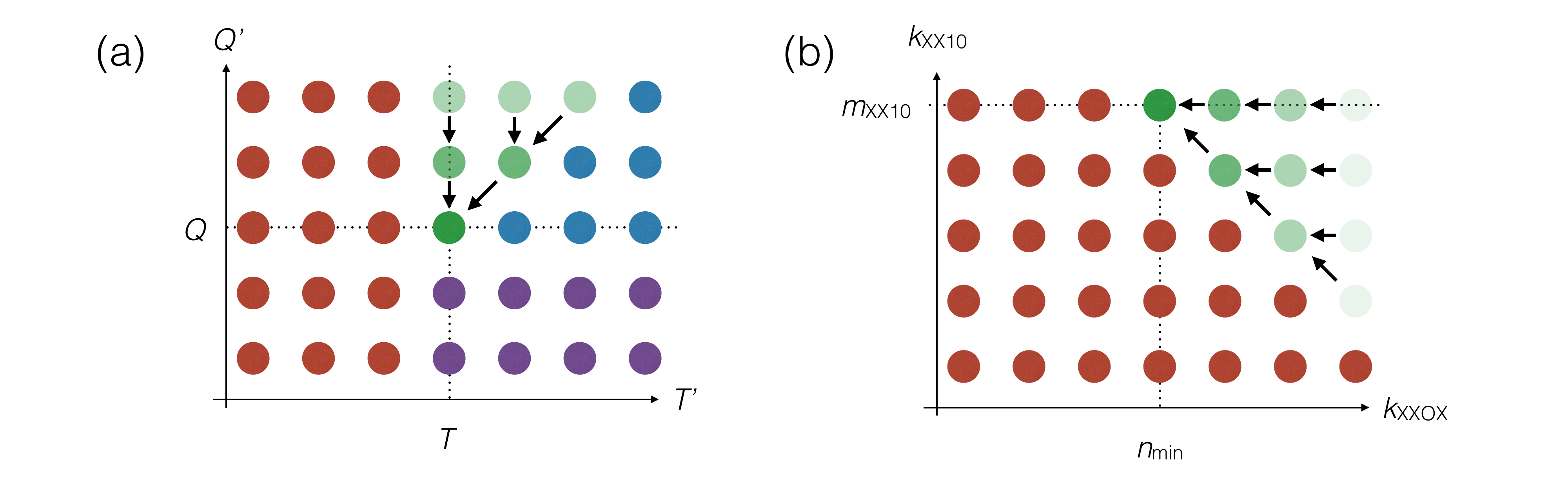}
    \caption{Optimal, suboptimal, and dysfunctional strategies.
    (a) Adversary strategies of a faulty sender $S$. 
    (b) Adversary strategies of a faulty receiver $R_0$.
    Filled green circle represents the incomplete optimal adversary strategy $\zeta_S$ in (a), $\zeta_R$ in (b). 
    Red, purple, and blue circles represent dysfunctional strategies. 
    Opaque green circles represent suboptimal adversary strategies. 
    Arrows depict increasing failure probability.}
    \label{fig:StrategyFailureFlow}
\end{figure}

(i)
If $T' < T$, then the strategy $\mathcal{\zeta}'_B$ is dysfunctional, $p_f = 0$, since
there are not enough indices  in  $\sigma_0$ and $\sigma_1$ to convince the receivers;
that is, the Length Condition of the Check Phase evaluates is violated, implying weak broadcast.
These strategies are depicted as red circles in Fig.~\ref{fig:StrategyFailureFlow}a.

(ii) If $T' \geq T$ and $Q' < Q$, then the strategy $\mathcal{\zeta}'_B$ is dysfunctional, since this excludes the possibility that $\rho_{01}$ contains enough inconsistent indices that compromise $R_0$.
These strategies are depicted as purple circles in Fig.~\ref{fig:StrategyFailureFlow}.

(iii) The failure probability of the strategy $\zeta'_B$ along the `line' $(T',Q') = (T+n,Q+n)$ decreases as $n\geq 0$ is increased.
I.e., along this line, depicted as the line of diagonal arrows in Fig.~\ref{fig:StrategyFailureFlow}a, the filled green circle depicting 
$n=0$ represents the optimal strategy.
The reason is that by increasing $T'$ from $T$ to $T+n$, the number of mixed outcomes in $\sigma_0$ for which $R_0$ needs to measure a `1' to cause failure is increased to $Q+n$, as demanded by the Consistency Condition of the Cross-Check Phase.
This reduces the corresponding failure probability from $(1/2)^Q$ to
$(1/2)^{Q+n}$.
This trend is depicted in Fig.~\ref{fig:StrategyFailureFlow}a by the diagonal arrows.

(iv) The strategy $\zeta'_B$ is dysfunctional `below the line' defined in (iii), see the blue circles in Fig.~\ref{fig:StrategyFailureFlow}a. 
There, even if all $Q'$ mixed outcomes included in $\sigma_0$ have `1' at $R_0$, that is insufficient to violate the Consistency Condition of the Cross-Check phase. 

(v) The failure probability of the strategies $\zeta'_B$ `above the line' defined in (iii), that is, strategies with $(T',Q') = (T+n,Q+n')$ (where $n\geq0$ and $n' \geq n$), decrease as $n'$ is increased.
Indeed, as $n'$ is increased, the check set size remains unchanged ($T+n$), but the number of `guessed' indices ($Q+n'$) is increased, and hence the failure probability $(1/2)^{Q+n'}$ decreases.
This trend is depicted in Fig.~\ref{fig:StrategyFailureFlow}a by the vertical arrows.

This concludes the proof that the strategy $\zeta = (T-Q,Q,0;0,0,m_{1100})$ is indeed optimal on its domain.

\subsection{\texorpdfstring{$R_0$ faulty}{R0 faulty}}
\label{app:r0faultyoptimality}

Here, we prove that the $R_0$ adversary strategy
$\mathcal{\zeta} = (0,m_\text{XX10},n_\text{min})$
defined above is optimal on its domain. 

(i) Increasing the first strategy parameter $k_{0011}$ from zero decreases the failure probability,
as it lowers the chance of the Consistency Condition of the Cross-check Phase evaluating as True.

(ii) Consider the alternative strategies $\zeta' = (0,k_\text{XX10},k_\text{XX0X})$, with $k_\text{XX10} = m_\text{XX10} - n$ and $k_\text{XX0X} = n_\text{min} + n'$, with  $n, n' \geq 0$.
The alternative strategies are dysfunctional ($p_f = 0$) `below the line' of the $(k_\text{XX10},k_\text{XX0X})$ plane defined by $n = n'$, i.e., for $n > n'$. 
This is because for such a strategy, the check set size $|\rho_{01}|$ is below the minimum size $T$ required to satisfy the Length Conditions. 
These dysfunctional strategies are shown as the red circles in Fig.~\ref{fig:StrategyFailureFlow}b.

(iii) The alternative strategies along the line $n=n'$, shown in Fig.~\ref{fig:StrategyFailureFlow}b as the line of diagonal arrows, have a decreasing failure probability as $n$ is increased. 
Along this line, the check set size $|\rho_{01}|$ is fixed to the minimum size $T$, but the number $n_\text{min} + n'$ of guessed indices increases, and hence the probability of satisfying the Consistency Condition of the Cross-check Phase decreases. 
This trend is depicted in Fig.~\ref{fig:StrategyFailureFlow}b by the diagonal arrows.

(iv) Consider the alternative strategies `above the line' of the $(k_\text{XX10},k_\text{XX0X})$ plane defined by $n = n'$, i.e., above the line of diagonal arrows in Fig.~\ref{fig:StrategyFailureFlow}b. 
These are the strategies of the form $\zeta = (0, m_\text{XX10}-n,n_\text{min}+n')$ with $n' \geq n$.
For a fixed $n$, the failure probability of these strategies decreases as $n'$ is increased.
This is because increasing $n'$ by 1 increases the check set size $|\rho_{01}|$ by 1, and hence the right hand side of the Consistency Condition of the Check Phase by 1, and at the same time, increases the minimum number of lucky indices needed to satisfy that Consistency Condition, leading to a decrease of the failure probability. 
This trend is depicted in Fig.~\ref{fig:StrategyFailureFlow}b by the horizontal arrows.

This concludes the proof that the strategy $\zeta = (0,m_\text{XX10},n_\text{min})$ is indeed optimal on its domain

\section{Security proof}
\label{app:securityproof}

In this Appendix, we derive asymptotic ($m \to \infty$) upper bounds for the failure probability for generic adversary strategies. 

\subsection{No faulty}

Without the loss of generality, we assume that $S$ sends the message $x_S =0$.

The protocol will achieve consensus, $y_0 = y_1 = x_S = 0$, if the number of 0011 outcomes in the event is at least $T$, cf. the exact failure-probability result of Eq.~\eqref{eq:pfnofaulty}.
The probability of achieving consensus, i.e., that a random event contains the outcome 0011 with frequency $m_{0011}$ is
\begin{eqnarray}
P(m_{0011}) = P\left( \left(\sum_{j=1}^m X_j\right) = m_{0011}\right), 
\end{eqnarray}
where $X_j$ are binary (Bernoulli) random variables, taking
values 0 or 1, with probability $p=1/3$ for the latter. 
This implies that 
\begin{eqnarray}
p_f(m) = 
P\left( \left(\sum_{j=1}^{m} X_j \right) \leq \mu m \right)
\end{eqnarray}
Recall that $\mu < 1/3$; hence, the failure probability $p_f$ can be upper bounded by the Chernoff bound as 
\begin{eqnarray}
p_f(m) \leq 
 \exp\left(
    - \frac 1 2 \frac m 3 (1-3\mu)^2
 \right).
\end{eqnarray}
That is, in the absence of an adversary, the failure probability of the Weak Broadcast protocol is subject to an upper bound that decreases exponentially as $m$ increases. 

Recall that the general form of the Chernoff bound used to derive this result is
\begin{equation}
\label{eq:chernoff1}
    P(X \leq (1-\delta) \bar{X}) \leq e^{- \bar{X} \delta^2/2},
    \mbox{ for }0 \leq \delta,
\end{equation}
where $\bar{X}$ is the expectation value of the sum variable $X$.

\subsection{\texorpdfstring{$S$ faulty}{S faulty}}

The security proof for the $S$ faulty adversary configuration will be based on the tight failure-probability upper bound calculation presented in App.~\ref{app:sfaultyfailureprobability}.

Our goal is here is to derive an exponentially decaying upper bound for the failure probability. 
To this end, we restrict the domain of the $S$ faulty adversary strategy $\zeta$, defined in \ref{app:sfaultyfailureprobability}.
The restricted strategy is denoted as $\zeta_r$.
In particular, we restrict the domain of $\zeta$ for `regular' or `typical' local count lists, where $\ell_1$ and $\ell_3$ are approximately equal to their expected values $m/3$.
Formally, we define $D(\zeta_r)$ as formed by those $(\ell_1,\ell_3)$ pairs where 
\begin{equation}
\label{eq:srestriction}
    \mu m < \ell_1, \ell_3 < 2m/3 - \mu m.
\end{equation}

When is $\zeta_r$ indeed a restriction of $\zeta$? 
Conditions \eqref{eq:cond1} and \eqref{eq:cond3} are automatically satisfied if Eq.~\eqref{eq:srestriction} holds.
However, Eq.~\eqref{eq:cond2} demands $Q \leq \ell_2$; this is also guaranteed if $\lambda \geq 1/2$.

Since $D(\zeta_r) \subset D(\zeta)$, and $D(\zeta)$ is optimal on its domain, the restricted strategy $\zeta_r$ is also optimal on its domain.

In the spirit of Eq.~\eqref{eq:psuppersimple}, the failure-probability upper bound can be expressed using the restricted strategy $\zeta_r$ as follows:
\begin{equation}
\label{eq:p_s_securityproof}
    p_{f}(m) \leq  
    \sum_{\ell \in D(\zeta_r)} 
    P_{\zeta_r}(\text{failure} | \ell)
    P(\ell) +
    \sum_{\ell \in \bar{D}(\zeta_r)} 
    P(\ell).
\end{equation}

It is straightforward to exponentially upper bound the first term using the fact that 
\begin{equation}
    P_{\zeta_r}(\text{failure}|\ell) = 2^{-Q} \approx 2^{-(1-\lambda)\mu m},
\end{equation} 
which is independent of $\ell$.
Therefore, we have
\begin{equation}
    \sum_{\ell \in D(\zeta_r)} 
    P_{\zeta_r}(\text{failure} | \ell)
    P(\ell) \approx
    2^{-(1-\lambda)\mu m}
    \sum_{\ell \in D(\zeta_r)} 
    P(\ell)
    \leq 
    2^{-(1-\lambda)\mu m},
\end{equation}
where we used $\sum_{\ell \in D(\zeta_r)}  P(\ell) \leq 1$.

To exponentially upper bound the second term of Eq.~\eqref{eq:p_s_securityproof}, we express it as
\begin{eqnarray}
\sum_{\ell \in \bar{D}(\zeta_r)} 
    P(\ell) &=& 
    P(
    (\ell_1 \leq \mu m) \mbox{ OR }
    (\ell_1 \geq 2m/3 - \mu m) \mbox{ OR }
    (\ell_3 \leq \mu m/3) \mbox{ OR }
    (\ell_3 \geq 2m/3 - \mu m)
    ) \nonumber
    \\
    &=&
    P(\ell_1 \leq \mu m) +
    P(\ell_1 \geq 2m/3 - \mu m) +
    P(\ell_3 \leq \mu m/3) +
    P(\ell_3 \geq 2m/3 - \mu m)
    \\
    &\leq&
    4 e^{-\frac 1 3 \frac m 3 (1-3\mu)^2}.
\end{eqnarray}
In the last step, we used another Chernoff bound (cf.~Eq.~\eqref{eq:chernoff1}): 
\begin{equation}
\label{eq:chernoff2}
    P(|X - \bar{X}| \geq \delta \cdot \bar{X} ) \leq 2 e^{- \bar{X} \delta^2/3},
    \mbox{ for }0 \leq \delta.
\end{equation}

\subsection{\texorpdfstring{$R_0$ faulty}{R0 faulty}}

Here, we describe the security proof for the $R_0$ faulty adversary configuration. This proof is based on the tight failure-probability upper bound calculation presented in App.~\ref{app:rfaultyfailureprobability}.

We restrict the domain of the $R_0$ faulty optimal incomplete adversary strategy $\zeta$, defined in \ref{app:rfaultystrategy}.
The restricted strategy is denoted as $\zeta_r$. 
In particular, we restrict the domain of $\zeta$ for `regular' local count lists, where $\ell_1$ and $\ell_2$ are close to their expected values, $m/3$ and $m/6$, respectively.
Formally, we define the domain $D(\zeta_r)$ of the restricted strategy as formed by those $(\ell_1,\ell_2)$ pairs which satisfy
\begin{eqnarray}
    \label{eq:rrestriction}
    \mu m &<  \ell_1  <& 2m/3 - \mu m, \\
    \mu m / 2 & <  \ell_2  < & 2m/6 - \mu m / 2.
\end{eqnarray}

Note that $\zeta_r$ is indeed a restriction of $\zeta$, since the upper bound in  Eq.~\eqref{eq:rrestriction} is a stricter condition than \eqref{eq:r0faultydomain}.
Furthermore, since $D(\zeta_r) \subset D(\zeta)$, and $\zeta$ is optimal on its domain, the restricted strategy $\zeta_r$ is also optimal on its domain. 

In the spirit of Eq.~\eqref{eq:psuppersimple}, the failure-probability upper bound can be expressed using the restricted strategy $\zeta_r$ as follows:
\begin{equation}
\label{eq:p_r_securityproof}
    p_{f}(m) \leq  
    \sum_{\ell \in D(\zeta_r)} 
    P_{\zeta_r}(\text{failure} | \ell)
    P(\ell) +
    \sum_{\ell \in \bar{D}(\zeta_r)} 
    P(\ell).
\end{equation}

To exponentially upper bound the second term of Eq.~\eqref{eq:p_r_securityproof}, we express it as 
\begin{eqnarray}
\sum_{\ell \in \bar{D}(\zeta_r)} 
    P(\ell) &=& 
    P(
    (\ell_1 \leq \mu m) \mbox{ OR }
    (\ell_1 \geq 2m/3 - \mu m) \mbox{ OR }
    (\ell_2 \leq \mu m/6) \mbox{ OR }
    (\ell_2 \geq 2m/6 - \mu m/2)
    ) \nonumber
    \\
    &=&
    P(\ell_1 \leq \mu m) +
    P(\ell_1 \geq 2m/3 - \mu m) +
    P(\ell_2 \leq \mu m/6) +
    P(\ell_2 \geq 2m/6 - \mu m/2)
    \\
    &\leq&
    2 e^{-\frac 1 3 \frac m 3 (1-3\mu)^2}
    +
    2 e^{-\frac 1 3 \frac m 6 (1-3\mu)^2}.
\end{eqnarray}
In the last step, we used the Chernoff bound in Eq.~\eqref{eq:chernoff2}.

Finally, we provide an upper bound for the first term of Eq.~\eqref{eq:p_r_securityproof}.
Within the restricted strategy domain $D(\zeta_r)$, the adversary $R_0$ has the greatest chance to cause failure if the number of `safe' indices, $\ell_2$, is maximal, $\ell_2 = \ell_{2,\text{max}} \equiv \lfloor 2m/6 - \mu m/2 \rfloor$. 
For a local count list containing $\ell_{2,\text{max}}$, Eq.~\eqref{eq:orangefailure} specifies $P_\zeta(\text{failure}|\ell_1,\ell_{2,\text{max}})$ as a sum of the upper part of a binomial distribution, between $k = T - Q + 1 - \ell_{2,\text{max}}$ to $k=T-\ell_{2,\text{max}}$.
In fact, the quantity $P_{\zeta_r}(\text{failure}|\ell_1,\ell_{2,\text{max}})$
is independent of $\ell_1$.

The Chernoff bound 
\begin{equation}
\label{eq:chernoff3}
    P(X \geq (1+\delta) \bar{X}) \leq e^{- \bar{X} \delta^2/3},
    \mbox{ for }0 \leq \delta,
\end{equation}
can be used to upper bound this sum, if the the latter is an upper tail sum, that is, the lower end of the $k$ sum is above the expectation value $\bar{X} = \frac 2 3 (T-\ell_{2,\text{max}})$ of $k$:
\begin{equation}
    {T-Q+1 - \ell_{2,\text{max}}} > \frac{2}{3} \left(T-\ell_{2,\text{max}}\right).
\end{equation}
The latter condition is readily transformed as
\begin{equation}
\label{eq:lambdacondition}
    \lambda > \frac{2+9\mu}{18\mu},
\end{equation}
where we disregarded ceiling and floor functions for simplicity.
Note that since $\lambda < 1$, the condition of Eq.~\eqref{eq:lambdacondition} can be satisfied only if 
\begin{equation}
\label{eq:mucondition}
    \mu > 2/9\approx 0.222.
\end{equation}

If the condition \eqref{eq:lambdacondition} holds, then the Chernoff bound of Eq.~\eqref{eq:chernoff3} is applied with the substitution
\begin{equation}
    \delta = \frac{2+ 9 \mu - 18 \lambda \mu}{6-27 \mu},
\end{equation}
yielding
\begin{eqnarray}
    \sum_{\ell \in D(\zeta_r)} P_{\zeta_r}(\text{failure}|\ell) P(\ell)
    &\leq& 
    \sum_{\ell \in D(\zeta_r)} P_{\zeta_r}(\text{failure}|\ell_1,\ell_{2,\text{max}}) P(\ell)
    \\
    &=&
    P_{\zeta_r}(\text{failure}|\ell_{2,\text{max}}) 
    \sum_{\ell \in D(\zeta_r)} P(\ell)
    \\
    &\leq& 
    P_{\zeta_r}(\text{failure}|\ell_{2,\text{max}}) 
    \\
    &\leq&
    e^{-\bar{X} \delta^2/3},
\end{eqnarray}
with
\begin{equation}
\label{eq:chernoff3applied}
    \bar{X} = \frac 2 3 (T-\ell_{2,\text{max}}) \approx \left(\frac 3 2 \mu - \frac 1 3 \right) m.
\end{equation}
Note that Eq.~\eqref{eq:mucondition} implies that the coefficient of $m$ on the right hand side is positive, i.e., the upper bound in Eq.~\eqref{eq:chernoff3applied} is exponentially decreasing with increasing $m$.

\section{Preparation and benchmarking of the four-qubit singlet
state on IBM Q}
\label{app:measurements}

This Appendix exhibits the data we obtained by conducting 
experiments on IBM Q quantum computer prototypes. 
We aimed these experiments to prepare the 
four-qubit singlet state $\ket{\psi}$ of Eq.~\eqref{eq:cabello}
using Circuit A of Fig.~\ref{fig:linear_circ}, 
and to quantify how close the experiment resembles
the preparation of an ideal four-qubit singlet state. 
Further details are in the table caption.

\begin{table}
\centering
\fontsize{8pt}{8pt}\selectfont
\begin{tabular}{| c | c | c | c | c | c | c| c | c |}
\hline
 backend & date & layout & $F_\text{c}$ & $F_\text{c}^\text{mitig}$ & $F_\text{q}$ & $F_\text{q}^\text{mitig}$ & $1-q \, [\%]$ & $1-q^\text{mitig}\, [\%]$ \\ 
 \hline
 \hline
 \multirow{8}{6em}{\verbperth} & 2023.07.13. & 1-3-0-5 & 0.814 & 0.976 & 0.621 & 0.813 & 81.9 & 97.98 \\  
 & 2023.07.13. & 3-5-1-4 & 0.817 & 0.941 & 0.691 & 0.862 & 84.5 & 97.29 \\
 & 2023.07.13. & 1-3-2-5 & 0.858 & \textcolor{red}{\bf{0.996}} & 0.696 & 0.873 & 86.0 & \textcolor{red}{\bf{99.96}} \\
 & 2023.07.13. & 3-5-1-6 & 0.840 & 0.943 & 0.712 & 0.852 & 86.2 & 96.77 \\
 & 2023.07.13. & 3-1-5-0 & 0.816 & 0.965 & 0.661 & 0.854 & 81.9 & 96.71 \\
 & 2023.07.13. & 5-3-4-1 & 0.820 & 0.938 & 0.582 & 0.716 & 82.6 & 94.41 \\
 & 2023.07.13. & 3-1-5-2 & 0.806 & 0.946 & 0.640 & 0.818 & 81.1 & 95.25 \\
 & 2023.07.13. & 5-3-6-1 & 0.837 & 0.933 & 0.587 & 0.694 & 84.0 & 93.65 \\
 \hline
 \multirow{8}{6em}{\verbjakarta} & 2023.07.13. & 1-3-0-5 & 0.829 & 0.941 & 0.721 & 0.892 & 83.3 & 94.53 \\
 & 2023.07.13. & 3-5-1-4 & 0.784 & 0.915 & 0.665 & 0.862 & 78.8 & 91.89 \\
 & 2023.07.13. & 1-3-2-5 & 0.849 & 0.951 & 0.740 & 0.888 & 85.3 & 95.56 \\
 & 2023.07.13. & 3-5-1-6 & 0.820 & 0.951 & 0.668 & 0.846 & 82.6 & 95.71 \\
 & 2023.07.13. & 3-1-5-0 & 0.819 & 0.932 & 0.687 & 0.854 & 82.0 & 93.31 \\
 & 2023.07.13. & 5-3-4-1 & 0.815 & 0.927 & 0.679 & 0.839 & 82.3 & 93.44 \\
 & 2023.07.13. & 3-1-5-2 & 0.851 & 0.954 & 0.729 & 0.873 & 85.5 & 95.73 \\
 & 2023.07.13. & 5-3-6-1 & 0.825 & 0.951 & 0.706 & 0.888 & 83.4 & 96.11 \\
  \hline
 \multirow{8}{6em}{\verbnairobi} & 2023.07.13. & 1-3-0-5 & 0.828 & 0.940 & 0.727 & 0.896 & 83.5 & 94.70 \\  
 & 2023.07.13. & 3-5-1-4 & 0.763 & 0.853 & 0.655 & 0.799 & 76.9 & 85.93 \\
 & 2023.07.13. & 1-3-2-5 & 0.823 & 0.937 & 0.704 & 0.877 & 82.9 & 94.28 \\
 & 2023.07.13. & 3-5-1-6 & 0.847 & 0.957 & 0.734 & 0.891 & 85.1 & 96.29 \\
 & 2023.07.13. & 3-1-5-0 & 0.818 & 0.925 & 0.708 & 0.868 & 82.2 & 92.84 \\
 & 2023.07.13. & 5-3-4-1 & 0.833 & 0.950 & 0.624 & 0.767 & 83.8 & 95.55 \\
 & 2023.07.13. & 3-1-5-2 & 0.791 & 0.895 & 0.667 & 0.828 & 79.5 & 89.84 \\
 & 2023.07.13. & 5-3-6-1 & 0.832 & 0.945 & 0.707 & 0.867 & 84.4 & 95.98 \\
   \hline
 \multirow{8}{6em}{\verblagos} & 2023.07.13. & 1-3-0-5 & 0.305 & 0.286 & 0.221 & 0.236 & 30.9 & 28.90 \\  
 & 2023.07.13. & 3-5-1-4 & 0.252 & 0.229  & 0.141 & 0.149 & 29.1 & 27.11 \\
 & 2023.07.13. & 1-3-2-5 & 0.309 & 0.296  & 0.258 & 0.275 & 30.9 & 29.67 \\
 & 2023.07.13. & 3-5-1-6 & 0.239 & 0.209  & 0.141 & 0.153 & 27.8 & 25.62 \\
 & 2023.07.13. & 3-1-5-0 & 0.398 & 0.427  & 0.286 & 0.321 & 82.2 & 88.54 \\
 & 2023.07.13. & 5-3-4-1 & 0.212 & 0.188  & 0.018 & 0.014 & 25.8 & 23.41 \\
 & 2023.07.13. & 3-1-5-2 & 0.393 & 0.415  & 0.259 & 0.284 & 84.6 & 89.77 \\
 & 2023.07.13. & 5-3-6-1 & 0.207 & 0.188  & 0.015 & 0.011 & 25.7 & 23.66 \\
 \hline
\multirow{4}{6em}{\verblima} & 2023.07.13. & 1-2-0-3 & 0.797 & 0.894 & 0.699 & 0.853 & 79.9 & 89.79 \\  
 & 2023.07.13. & 2-1-3-0 & 0.806 & 0.906  & 0.681 & 0.830 & 80.7 & 90.64 \\
 & 2023.07.13. & 2-3-1-4 & 0.809 & 0.913  & 0.706 & 0.868 & 81.1 & 91.57 \\
 & 2023.07.13. & 3-2-4-1 & 0.809 & 0.916  & 0.672 & 0.829 & 81.1 & 91.69 \\
 \hline
\multirow{4}{6em}{\verbmanila} & 2023.07.15. & 1-2-0-3 & 0.849 & 0.961 & 0.715 & 0.872 & 85.0 & 96.27 \\  
 & 2023.07.15. & 2-1-3-0 & 0.847 & 0.949  & 0.716 & 0.859 & 85.2 & 95.52 \\
 & 2023.07.15. & 2-3-1-4 & 0.847 & 0.945  & 0.711 & 0.844 & 84.9 & 94.68 \\
 & 2023.07.15. & 3-2-4-1 & \textcolor{red}{\bf{0.875}} & 0.977  & \textcolor{red}{\bf{0.757}} & 0.895 & \textcolor{red}{\bf{87.8}} & 97.99 \\
 \hline
\multirow{4}{6em}{\verbbelem} & 2023.07.28. & 1-2-0-3 & 0.636 & 0.758 & 0.412 & 0.566 & 63.9 & 76.07 \\  
 & 2023.07.28. & 2-1-3-0 & 0.687 & 0.852  & 0.467 & 0.685 & 69.4 & 85.74 \\
 & 2023.07.28. & 2-3-1-4 & 0.653 & 0.798  & 0.468 & 0.677 & 66.0 & 80.88 \\
 & 2023.07.28. & 3-2-4-1 & 0.717 & 0.891  & 0.507 & 0.737 & 73.0 & 90.46 \\
\hline
\multirow{4}{6em}{\verbquito} & 2023.07.29. & 1-3-0-4 & 0.793 & 0.950 & 0.661 & 0.893 & 80.1 & 95.86 \\  
 & 2023.07.29. & 3-1-4-0 & 0.768 & 0.920  & 0.650 & 0.883 & 77.4 & 93.09 \\
 & 2023.07.29. & 1-3-2-4 & 0.779 & 0.964  & 0.656 & \textcolor{red}{\bf{0.931}} & 78.3 & 97.07 \\
 & 2023.07.29. & 3-1-4-2 & 0.756 & 0.948  & 0.630 & 0.902 & 76.8 & 96.42 \\
\hline
\end{tabular}
\caption{Preparation and benchmarking of the four-qubit singlet
state on IBM Q prototype quantum computers. 
We prepared the state $\ket{\psi}$ of Eq.~\eqref{eq:cabello}
using the linear circuit shown in Fig.~\ref{fig:linear_circ}.
The first column indicates the quantum computer (`backend') used;
the second column shows the date of the measurement; the third column represents the layout mapping the
physical qubits of the backend to the virtual qubits of the linear circuit.
The fourth (fifth) column, $F_c$ ($F_c^\text{mitig}$), shows 
the classical fidelity, Eq.~\eqref{eq:classicalfidelity},
of the ideal distribution (Fig.~\ref{fig:Cabstate_distribution}c)
and the distribution generated by the backend,
without (with) readout error mitigation. 
The sixth (seventh) column,
$F_q$ ($F_q^\text{mitig}$), shows 
the quantum state fidelity, Eq.~\eqref{eq:quantumfidelity},
of $\ket{\psi}$ 
and the output density matrix of 
the linear circuit, as inferred from quantum state tomography
without (with) readout error mitigation.
Measured `noise strength' values $q$ (as defined in Sec.~\eqref{subsec:robustness}) without (with) readout error mitigation are indicated in column 8 (9). Best values are colored red.
\label{tab:ibmq_results}}
\end{table}

\end{document}